\documentclass{aastex62}

\usepackage[figuresright]{rotating}
\usepackage{multirow}

\graphicspath{{./}{figures/}}

\received{2019 September 3}
\revised{2019 November 4}
\accepted{2019 November 22}
\submitjournal{ApJ}

\shorttitle{Rare Low Mass LMC Wolf-Rayet Star}
\shortauthors{Margon et al.}

\def\CII{\hbox{C~$\scriptstyle\rm II$}}
\def\CIII{\hbox{C~$\scriptstyle\rm III$}}
\def\CIV{\hbox{C~$\scriptstyle\rm IV$}}
\def\HeI{\hbox{He~$\scriptstyle\rm I$}}
\def\HeII{\hbox{He~$\scriptstyle\rm II$}}
\def\NII{\hbox{N~$\scriptstyle\rm II$}}
\def\fNII{\hbox{[N~$\scriptstyle\rm II$]}}

\def\NeII{\hbox{Ne~$\scriptstyle\rm II$}}

\def\OII{\hbox{O~$\scriptstyle\rm II$}}
\def\MgII{\hbox{Mg~$\scriptstyle\rm II$}}
\def\OIII{\hbox{O~$\scriptstyle\rm III$}}
\def\OI{\hbox{O~$\scriptstyle\rm I$}}
\def\fOI{\hbox{[O~$\scriptstyle\rm I$]}}
\def\fOII{\hbox{[O~$\scriptstyle\rm II$]}}
\def\fSII{\hbox{[S~$\scriptstyle\rm II$]}}
\def\SiII{\hbox{Si~$\scriptstyle\rm II$}}
\def\SiIII{\hbox{Si~$\scriptstyle\rm III$}}
\def\CPD{\hbox{CPD~$-56^\circ\,8032$}}
\def\fOIII{\hbox{[O~$\scriptstyle\rm III$]}}

\def\Gaia{\hbox{\it{Gaia}}}
\def\Galex{\hbox{\it{GALEX}}}
\def\WISE{\hbox{\it{WISE}}}
\def\kms{\hbox{km~s$^{-1}$}}

\begin{document}

\title{Discovery of a Rare Late-Type, Low-Mass Wolf-Rayet Star in the LMC}

\author{Bruce Margon}
\affiliation{Department of Astronomy \& Astrophysics, University of California, Santa Cruz, 1156 High St, Santa Cruz, CA 95064, margon@ucsc.edu}

\author{Catherine Manea}
\affiliation{Department of Astronomy \& Astrophysics, University of California, Santa Cruz, 1156 High St, Santa Cruz, CA 95064, margon@ucsc.edu}
\affiliation{Current Address: Department of Astronomy, University of Texas at Austin, Austin, TX 78712}

\author{Robert Williams}
\affiliation{Department of Astronomy \& Astrophysics, University of California, Santa Cruz, 1156 High St, Santa Cruz, CA 95064, margon@ucsc.edu}
\affiliation{Space Telescope Science Institute, 3700 San Martin Drive, Baltimore, MD 21218}

\author{Howard E. Bond}
\affiliation{Space Telescope Science Institute, 3700 San Martin Drive, Baltimore, MD 21218}
\affiliation{Department of Astronomy \& Astrophysics, Pennsylvania State University, University Park, PA 16802}

\author{J. Xavier Prochaska}
\affiliation{Department of Astronomy \& Astrophysics, University of California, Santa Cruz, 1156 High St, Santa Cruz, CA 95064, margon@ucsc.edu}
\affiliation{Kavli Institute for the Physics and Mathematics of the Universe, University of Tokyo, 5-1-5 Kashiwanoha, Kashiwa, Chiba 277-8583, Japan}

\author{Micha\l~K. Szyma\'{n}ski}
\affiliation{Warsaw University Observatory, Al. Ujazdowskie 4, PL-00-478 Warszawa, Poland}

\author{Nidia Morrell}
\affiliation{Las Campanas Observatory, Carnegie Observatories, Casilla 601, La Serena, Chile}

\begin{abstract}

We report the serendipitous discovery of an object, UVQS J060819.93$-$715737.4, with a spectrum dominated by extremely intense, narrow \CII{} emission lines.  The spectrum is similar to those of the very rare, late-type [WC11] low-mass Wolf-Rayet stars.  Despite the recognition of these stars as a distinct class decades ago, there remains barely a handful of Galactic members, all of which are also planetary-nebula central stars.  Although no obvious surrounding nebulosity is present in J0608, \fOII{}, \fNII{}, and \fSII{} emission suggest the presence of an inconspicuous, low-excitation nebula. There is low-amplitude incoherent photometric variability on timescales of days to years, as well as numerous prominent P~Cygni profiles, implying mass loss. There are indications of a binary companion.  The star is located on the outskirts of the LMC, and the observed radial velocity ($\sim \!+250$~\kms{}) and proper motion strongly suggest membership. If indeed an LMC member, this is the first extragalactic late [WC] star, and the first with an accurately determined luminosity, as the Galactic examples are too distant for precise parallax determinations. A high-quality, broad-coverage spectrum of the prototype of the late [WC] class, \CPD{}, is also presented.  We discuss different excitation mechanisms capable of producing the great strength of the \CII{} emission.  Numerous autoionizing levels of \CII{} are definitely populated by processes other than dielectronic recombination. Despite the spectacular emission spectra, observational selection makes objects such as these difficult to discover. Members of the [WC11] class may in fact be considerably more common than the handful of previously known late [WC] stars.

\end{abstract}


\keywords{stars: Wolf-Rayet --- 
planetary nebula: individual --- radiation mechanisms: general --- stars:  chemically peculiar}

\section{Introduction}\label{sec:introduction}

During the UV-Bright Quasar Survey (UVQS) for intermediate-redshift QSOs bright in the far-UV (Monroe et al.\ 2016), several unusual stars also met the Survey's \Galex\/ (ultraviolet) and \WISE\/ (infrared) color-selection criteria, and were therefore observed spectroscopically (e.g., Margon et al.\ 2016). Here we report on another such unusual object, UVQS J060819.93$-$715737.4 ($l=283^\circ$, $b=-29^\circ$, hereafter J0608), which has numerous intense emission lines of \CII{}. This star is likely a new member of the extremely rare class of cool, late-type low-mass Wolf-Rayet (W-R) stars, a known but small collection of stellar objects whose spectra are 
dominated by extremely strong \CII{} emission. Over half a century 
ago, one of us (HEB), while examining objective-prism plates from the Curtis Schmidt 
telescope at the Cerro Tololo Inter-American Observatory, discovered that CPD\,$-56^\circ$\,8032 (= Hen\,3-1333, = 
V837~Ara) is an emission-line object having \CII{} $\lambda$4267 and $\lambda$6580 as its strongest features (Bidelman, MacConnell, \& Bond 1968). It was suggested that the star could be an unusually cool WC-type W-R star. More detailed studies of the spectrum of this object are presented by Cowley \& Hiltner (1969), Thackeray (1977), and numerous later authors.

It has been known for over a century that some central stars of planetary nebulae (PNe) have spectra that resemble those of classical W-R stars. An early example was BD\,$+30^\circ$\,3639 (= Campbell's Hydrogen-Envelope Star; Campbell 
1893), whose emission-line spectrum had been discovered on a Harvard objective-prism plate by Williamina Fleming, and announced by Pickering (1891). Campbell showed that the object is surrounded by a resolved, compact PN.

The modern understanding of classical W-R stars is that they are massive young stars with dense stellar winds. The W-R central stars of PNe, however, are a separate class of considerably less massive, older, and highly evolved objects, which also have dense outflows---as recognized as long ago as Smith \& Aller (1971) in their discussion of Campbell's Star, and in earlier papers which they reference. To recognize this fundamental difference, van der Hucht et al.\ (1981) suggested 
that the spectral types for low-mass W-R stars be enclosed in square brackets. Thus, for example, the carbon-rich, low-excitation spectrum of BD\,$+30^\circ$\,3639 is classified [WC9].

Webster \& Glass (1974, hereafter WG74) suggested that CPD\,$-56^\circ$\,8032, 
along with PN~M\,4-18 (=~PK~146+07~1), Hen\,2-113, and possibly V348~Sgr, could be considered to be 
members of a class of objects with spectra dominated by strong \CII{} emission. They appear to form an extension of the PN nuclei with WC spectra to 
even lower excitation levels than objects like BD\,$+30^\circ$\,3639. WG74 also 
pointed out that all four objects have strong infrared (IR) emission, implying a 
substantial dust component. Radial velocities, available for three of the objects, were all relatively large, indicating membership in an old disk or Galactic-bulge population. In the years since the WG74 paper, although many early [WC] stars have been found among PNe nuclei, only a few with late-[WC] spectra have been identified (G\'{o}rny 2001; DePew et al.\ 2011).

The organization of this paper is as follows. We discuss our observations of the newly discovered object in \S2, and the detailed nature of the spectrum in \S3. The spectral-energy distribution is examined in \S4, and inferred evolutionary status of the star in \S5. Comments are made on the possible relationship to PNe in \S6 and inferences from photometry in \S7. The interesting absorption-line system is discussed in \S8, a detailed description of the physics of the \CII{} emission appears in \S9, and conclusions are summarized in \S10.

\section{Observations}\label{sec:observations}

J0608 is noted by Monroe et al.\ (2016) as a Galactic source (i.e., not an AGN), without further comment. Based on good positional coincidence, we conclude that the star also appears in the {\it IRAS\/} (IRAS 06091$-$7157) and {\it AKARI\/} catalogs, as well as in \WISE\/ and \Galex\/ (required for UVQS selection), again without comment. The bright IR fluxes suggest a dusty local environment. 

The UVQS spectrum of J0608 was obtained on UT 2015 February 14 with the Boller and Chivens spectrograph at the 2.5m du Pont telescope of the Las Campanas Observatory. A 180~s exposure with the 600 $l$~mm$^{-1}$ grating blazed at 5000 \AA\ and a $2\arcsec$ slit yielded $\sim$4 \AA~spectral resolution. The data were sky-subtracted and extracted with standard, optimal techniques. Although we suspect the presence of underlying nebulosity (see \S6), as we have no knowledge of the morphology of any extended component (other than it is quite compact), we cannot quantitatively assess its contribution to the integrated spectrum. If low-level nebulosity exists, it would have been subtracted as if it were sky, but this is expected to be a small effect. Approximate flux calibration was obtained via observations of spectrophotometric standard stars. A finding chart and FITS format file of the spectrum are available in the MAST archive of UVQS data,\footnote{\url{https://archive.stsci.edu/prepds/uvqs/target_pages/j060819.93-715737.4.html}} although the suggested line identifications and derived redshift there presume a QSO, and thus should be ignored.

The observed flux indicates $V\simeq15.2$ at the time of that exposure, although uncertain light losses at the slit require that this estimate be used with caution. At $b=-29^\circ$, the anticipated interstellar extinction is modest; the \Galex\/ catalog estimates $E(B-V)=0.09$. Photometry from SkyMapper DR1 (Wolf et al.\ 2018) yields $r=15.63$, $(g-r)=-0.14$, $(u-g)=-0.16$, $(r-i)=-0.19$.

\begin{sidewaysfigure}
    \includegraphics[scale=.9]{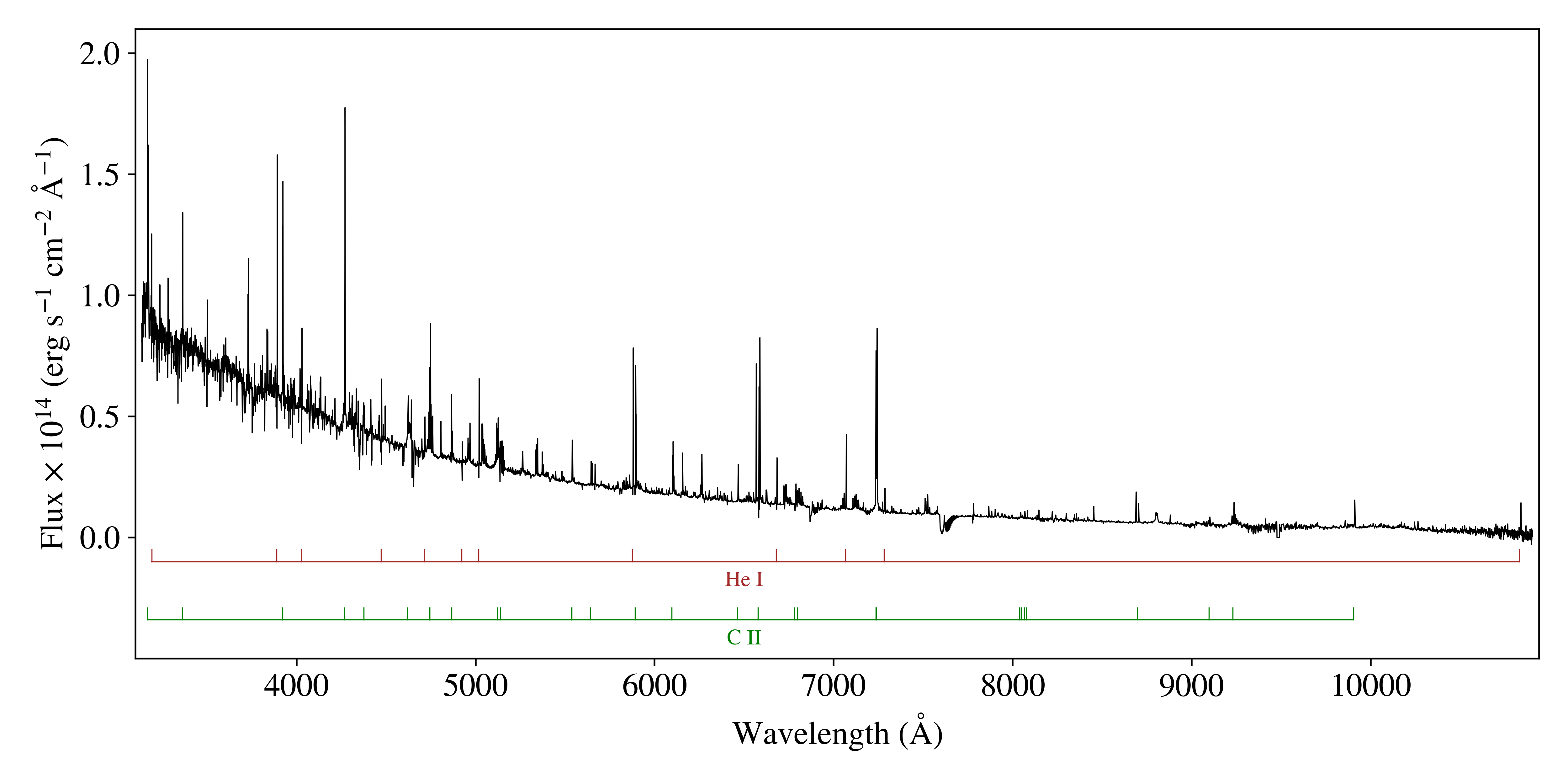}
    \caption{A Magellan MagE spectrum of J0608, obtained in 2019 May. The spectrum is dominated by \CII{} $\lambda$4267 and $\lambda$7231 emission, but numerous other \CII{} and \HeI{} emission lines are also indicated. Note the marked blue excess in the continuum, the P~Cygni absorption profiles on many of the emission lines, and the marked increase in the occurrence of absorption lines shortward of $\sim$5200~\AA. Although not obvious in this figure, at the two extreme ends of the spectrum prominent emission lines of the \CII{} $\lambda\lambda 3165, 3167$ doublet and \HeI{} $\lambda10830$ are also well detected. Suggested line identifications are given in Table 1.
    \label{fig:f1}}
\end{sidewaysfigure}

The unusual nature of J0608 in the UVQS spectrum prompted us to seek considerably higher signal-to-noise spectra with better resolution, using the Magellan Echellette Spectrograph (MagE) on the Baade Magellan telescope (Marshall et al.\ 2008). Multiple exposures totaling 4,200 s were obtained on 2017 December 30, using a $1\arcsec$ slit which provides $\sim$1 \AA~spectral resolution. A second spectrum was obtained on 2019 May 3 with a similar configuration. Flux calibration was obtained via observations of multiple spectrophotometric standard stars, but must be regarded as approximate.  The two spectra are qualitatively similar; the 2019 spectrum is displayed in Figure~\ref{fig:f1}. The changes in intensity between the two spectra are consistent with the differences in the signal-to-noise ratios and standard star calibrations, but do not rule out changes in the physical conditions in the wind between the two epochs of observation.

\section{The Spectrum of J0608} \label{sec:spectrum}

Several unusual features of J0608 are readily apparent in Figure~\ref{fig:f1}. The strongest emission lines prove to be due to \CII{}, perhaps not immediately familiar to many stellar spectroscopists, with $\lambda$7236, $\lambda$6580, $\lambda$4267 dominating the entire spectrum. Other prominent emission lines include narrow H$\alpha$ and H$\beta$ (but not the higher-order Balmer lines), as well as numerous \HeI{} features, \OI{} $\lambda$7773, $\lambda$8446, and prominent nebular lines, including \fOII{}  $\lambda\lambda$3726, 3729, [\NII{}] $\lambda$6548, $\lambda$6583, and \fSII{} $\lambda$6717, $\lambda$6731. There are also numerous absorption lines; many but by no means all have P~Cygni profiles. A list of suggested line identifications, together with equivalent widths and relative intensities in the two spectra, is given in Table~1. For lines which are purely in emission or absorption, wavelengths and equivalent widths are derived using the ALIS software package (Cooke et al.\ 2014), which includes a fit to the local continuum.  This package employs a $\chi^2$ minimization procedure to minimize the residuals between the data and a user-specified model, weighted by the inverse variance of the data\footnote{\hyperlink{https://github.com/rcooke-ast/ALIS}{https://github.com/rcooke-ast/ALIS}}. For lines with P~Cyg profiles, the quoted wavelength is where the emission and absorption components transition into each other at the level of the continuum. As this object is known to be photometrically variable (see \S7), and we have no accurate photometry contemporaneous with either of the two spectra, any apparent differences in flux between the two data sets in Table~1 should be regarded with caution.

It is clear from Table~1 that the emission lines display a substantial and consistent significantly positive radial velocity. For example the mean of the 110 strongest identified emission lines in the 2017 spectrum is $+247.1 \pm 1.4$ \kms{}, where the quoted uncertainty is the mean standard deviation.  There is a small but formally statistically significant difference in the absorption-line radial velocity (excluding P~Cyg absorptions), measured from 39 lines, of $+217.6 \pm 4.3$ \kms{}.  An obvious blue continuum excess is also evident, consistent with the NUV and FUV detections by \Galex\/, as well as the marked UV-excess in the SkyMapper photometry.

The precise sub-classification of the very latest-type [WC] stars is somewhat author-dependent, hardly surprising given the paucity of class members. The topic is well-discussed by Crowther, De Marco, \& Barlow (1998), de Araujo et al.\ (2002), Acker \& Neiner (2003), Weidmann \& Gamen (2011), and references therein. Today the latest, lowest-excitation end of this sequence is typically although not universally denoted as [WC11], and while we recognize that any sub-classification of J0608 will not be embraced unanimously, for the sake of brevity and specificity we will henceforth also refer to this object as [WC11]. The lack of both \CIV{} and \HeII{} emission in J0608 comports well with the [WC11] class definition of Crowther et al. (1998). J0608 may well be one of the lowest-excitation [WC11] examples known, as our observed \CII{} $\lambda$4267/\CIII{} $\lambda$5696 ratio is more extreme than other published examples of late-type [WC] stars, including the prototype \CPD{} (de Araujo et al.\ 2002), and more recently discovered low-excitation examples (Pe\~{n}a 2005). 

As we are aware of no published, modern, broad-coverage digital spectrum of the prototype \CPD{}, for comparison we present one in Figure~2, kindly obtained by P.~Massey, also with MagE at the Magellan telescope. J0608 clearly has a strong spectral resemblance, although \CPD{} lacks the extreme blue continuum excess.

\begin{figure}
\centering
\includegraphics[scale=.7]{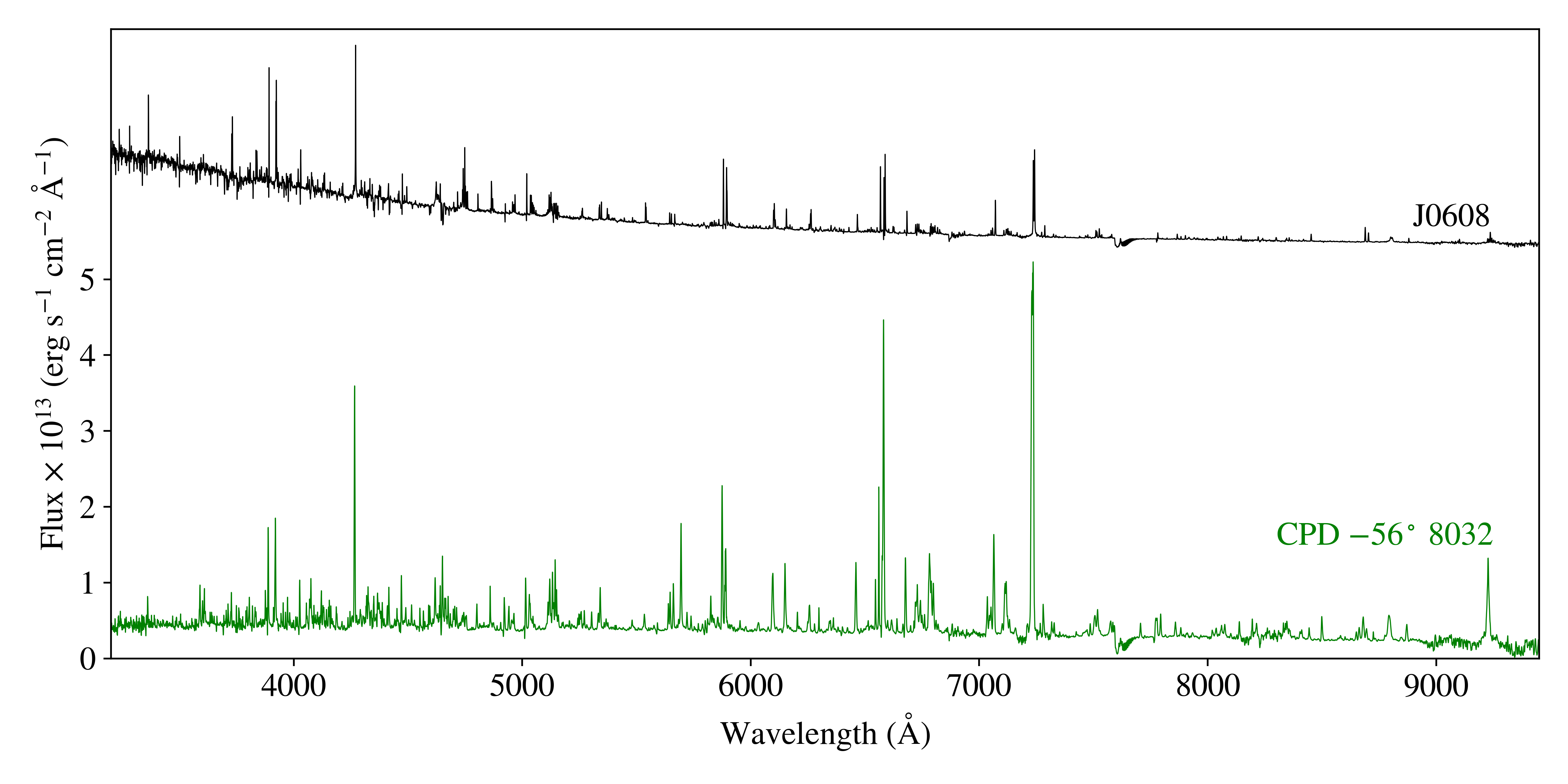}
\caption{{\it Lower:} Spectrum of \CPD{}, the [WC11] prototype, also obtained with Magellan and MagE on 4 February 2018, courtesy of P.~Massey. {\it Upper:} The spectrum of J0608, reproduced from Fig.~1. This J0608 spectrum has been multiplied by a constant and then additively shifted upwards for clarity of display; absolute flux values can be obtained from Fig.~1. The similarity of \CPD{} to J0608 is evident, although the steep blue continuum of the latter star is missing, possibly due to circumstellar dust (Aitken et al. 1980, Kameswara Rao et al. 1990).}
\end{figure}

It has been suggested by Crowther et al.\ (1998) that V348 Sgr be removed from the already small class, due to the absence of \CIII{} $\lambda$5696, which is indeed also weak, although definitely present, in emission in J0608. However, V348~Sgr is widely acknowledged to be an R CrB star, with violent and repeated photometric variability, behavior known to be absent in J0608 (cf.\ \S7). Similar discussions appear in the literature regarding HV 2671 in the LMC (De Marco et al.\ 2002), and the more general question of a possible link between [WC11] and R~CrB stars is intriguing but unsettled (Clayton et al.\ 2011).

\begin{figure}
\centering
\includegraphics[scale=.65]{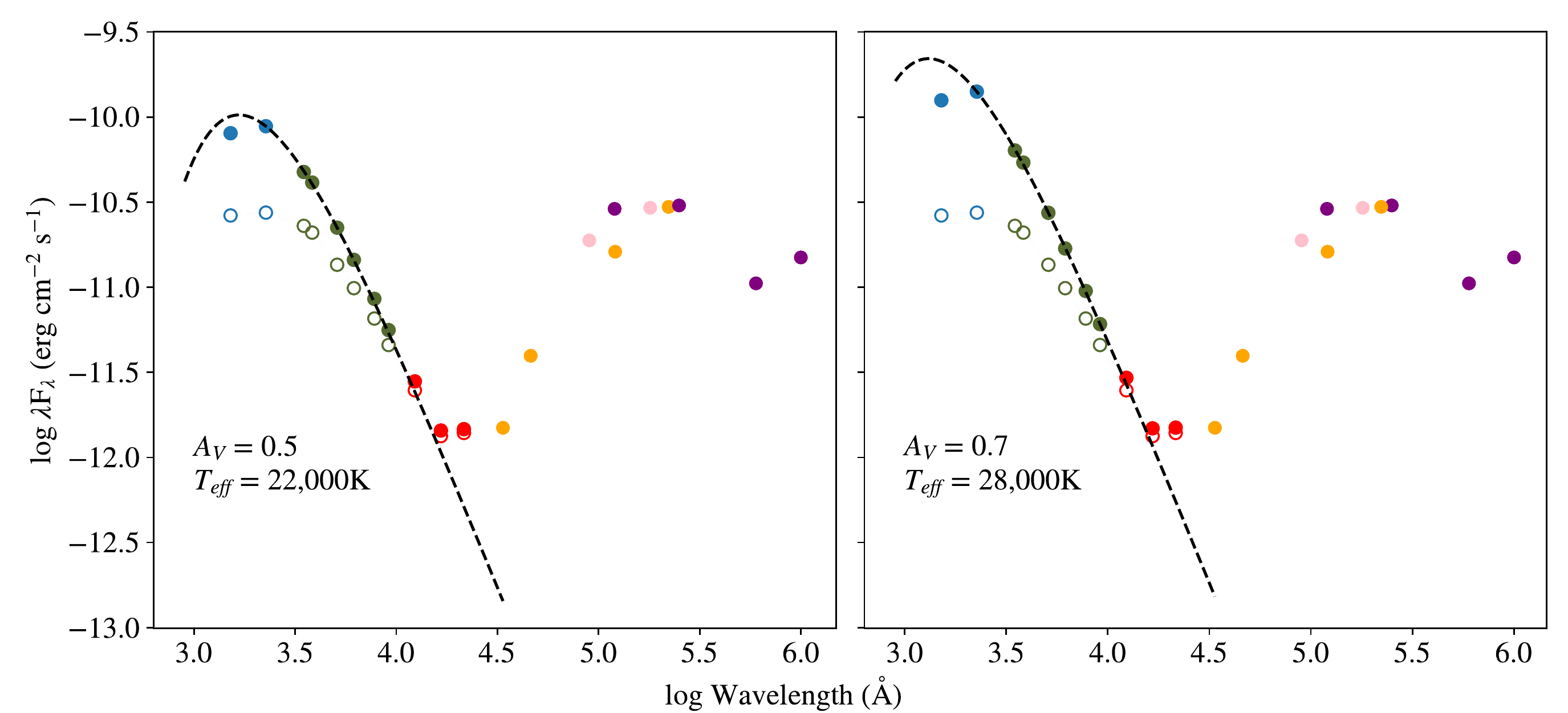}
\caption{The spectral-energy distribution of J0608, utilizing data from  \Galex\/ ({\it blue points}), SkyMapper ({\it green points}), 2MASS ({\it red points}), \WISE\/ ({\it orange points}), {\it AKARI\/} ({\it pink points}), and {\it IRAS\/} ({\it purple points}). Two extinction-corrected cases are considered, with parameters described in the text. {\it Open circles:} raw photometry; {\it filled circles}: extinction-corrected data. Plausible blackbody fits to the UV and visible data ({\it broken lines}) have $T_{\rm eff}$=22,000~K ({\it left panel}) and 28,000~K ({\it right panel}), but these fits are not unique. The infrared data are clearly in excess of the flux from the hot source, and this excess is likely due to circumstellar dust, ubiquitous in late [WC] stars. Note that none of the data sources are contemporaneous.}
\end{figure}

\section{The Spectral-Energy Distribution}

As substantial photometry is available for J0608 from the UV through the IR, the overall spectral-energy distribution (SED) of the object is clearly of interest, and is displayed in Figure~3, with two different extinction cases considered. In addition to the previously noted UV excess, there is a distinct IR excess. Because the object is known to be at least a low-amplitude photometric variable, and these data are not contemporaneous, caution must be taken interpreting the photometry. 

The rising blue continuum and the prominent \Galex\/ detections indicate the presence of a hot source. As a very crude temperature indicator, we fit blackbodies to the UV and visible photometry. For this purpose, the blue and UV observed fluxes must be corrected for extinction. While the interstellar extinction in this direction is well constrained by a variety of data, and is small, the strong 2MASS and \WISE\/ detections in the IR imply that circumstellar absorption due to dust is also important, and quite possibly the dominant component of the total extinction.

We use the Fitzpatrick (1999) and Sun et al.\ (2018) models to correct the optical and UV data, respectively, and find that there is a range of $A_V$ values that provide reasonable fits to the data of Figure~3. We show two examples, with ($A_V$, $T_{\rm eff}$) = (0.5, 22,000~K) and (0.7, 28,000~K). We caution that these models are by no means unique solutions, and are also sensitive to the assumed value of $R_V$, the optical total-to-selective extinction ratio. The familiar value of $R_V=3.1$, usually taken for the interstellar medium and assumed here, may not necessarily apply, due to substantial circumstellar extinction. Not surprisingly, $T_{\rm eff}$ is also extremely sensitive to the shortest-wavelength point, the \Galex\/ FUV flux.

The bright IR components are certainly compatible with the marked IR excesses known to characterize the dusty late [WC] stars (WG74). The longest-wavelength points are clearly far in excess of the hot blackbody that creates the UV and optical emission. The mid- to far-infrared points cannot be fit with a single blackbody, but require multiple components in the $T_{\rm eff}$ = 100 -- 500~K range.  Again, these parameters should not be regarded as unique.


\section{LMC Membership, Host Population, and Evolutionary State} 
\label{sec:distandlum}

J0608 is located $\sim\!5^\circ$ from the center of the LMC, but its +247~\kms\
emission-line radial velocity is nevertheless consistent with LMC membership (e.g., van der
Marel et al.\ 2002). Nidever et al.\ (2019) and others have shown that LMC
members extend well beyond the position of J0608. An alternative distance
scenario placing J0608 in our Galaxy would require halo membership, in order to
be compatible with the high radial velocity and moderately high Galactic
latitude. However, the known Galactic late [WC] objects are fairly obviously not
members of an especially old population, as they tend to be found at
low Galactic latitudes.

\begin{figure}[b]
\centering
\includegraphics[scale=.4]{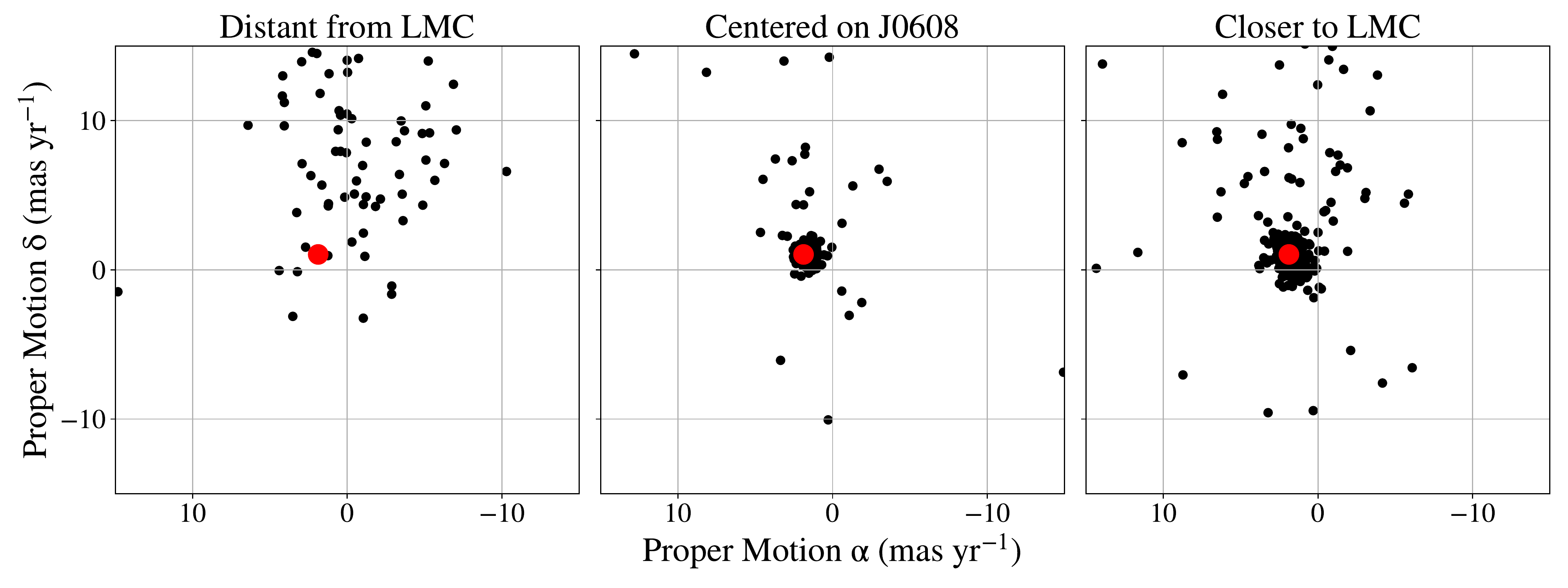}
\caption{\Gaia{} DR2 proper motions near J0608 and environs. {\it Center panel}: A square of $4\arcmin$ centered on J0608 ({\it red dot}). {\it Left panel}: $4\arcmin$ field $\sim$20$^\circ$ from the LMC but at the same galactic latitude as J0608, presumably all halo stars; note the large scatter in the halo-star proper motions. {\it Right panel}: $4\arcmin$ field $\sim$3$^\circ$ from the LMC center, dominated by LMC members. The observed motion of J0608 is superposed on the left and right panels. The motion of J0608 strongly implies LMC membership.\label{fig:f4}}
\end{figure}

The recent \Gaia\/ Data Release~2 (DR2) provides measurements of the parallax, $\pi$,
and proper motion, $\mu$, of J0608 and neighboring stars (Lindegren et al.\ 2016;
Gaia Collaboration et~al.\ 2018), allowing an opportunity to examine the issue
of LMC membership. The DR2 parallax is $\pi=0.0035\pm0.0321$~mas, statistically
consistent with an LMC distance ($\pi\simeq0.020$~mas), but also not ruling out membership in the distant Galactic halo. The DR2 proper motions of J0608 are
$\mu_\alpha = 1.875 \pm 0.059$ mas\,yr$^{-1}$ and $\mu_\delta = 1.028 \pm 0.068$
mas\,yr$^{-1}$. These values are nicely consistent with \Gaia\/ data for LMC
members (e.g., Helmi et al.\ 2018; Vasiliev 2018), whereas foreground
Galactic-halo stars in this direction have a large dispersion in $\mu$. These
points are illustrated in Figure~\ref{fig:f4}, in which \Gaia\/ proper motions for all stars within $2'$
of the location of J0608 are shown, as well as for two other fields of the same size, one
closer to the center of the LMC, and one considerably farther, but at the same
Galactic latitude. The LMC members are tightly clustered in proper-motion space, and
J0608's proper motion lies well within this LMC distribution.

LMC membership for J0608 is nevertheless a bit surprising for two reasons. The LMC is well-studied and known to contain only early-type WC stars (e.g., Breysacher et al.\ 1999), a conclusion upheld even by the most recent surveys (Massey et al.\ 2017). Even the known LMC [WC] stars are only found to have spectral types of [WC8] or earlier (Pe\~na et al.\ 1997; Crowther 2008). Moreover, the implied absolute luminosity of J0608 at the LMC distance, $M_V\simeq-3$, would exceed that normally quoted for the known Galactic [WC11] stars; however, luminosity estimates for the
[WC11] subgroup are admittedly quite uncertain, as none are close enough for an accurate \Gaia\/ DR2 parallax. The known early [WC] stars in the LMC, where distance uncertainty is not an issue, are also all less luminous than inferred for J0608 if it is a member, despite their earlier spectral types (Pe\~na et al.\ 1997). However, LMC membership may shed light on the marked weakness of \ion{C}{3} $\lambda$5696 emission in J0608: Torres et al.\ (1986) point out that all Galactic WC's have \ion{C}{3} $\lambda$5696 emission, but many in the LMC do not.

\begin{figure}
\centering
\includegraphics[scale=.6]{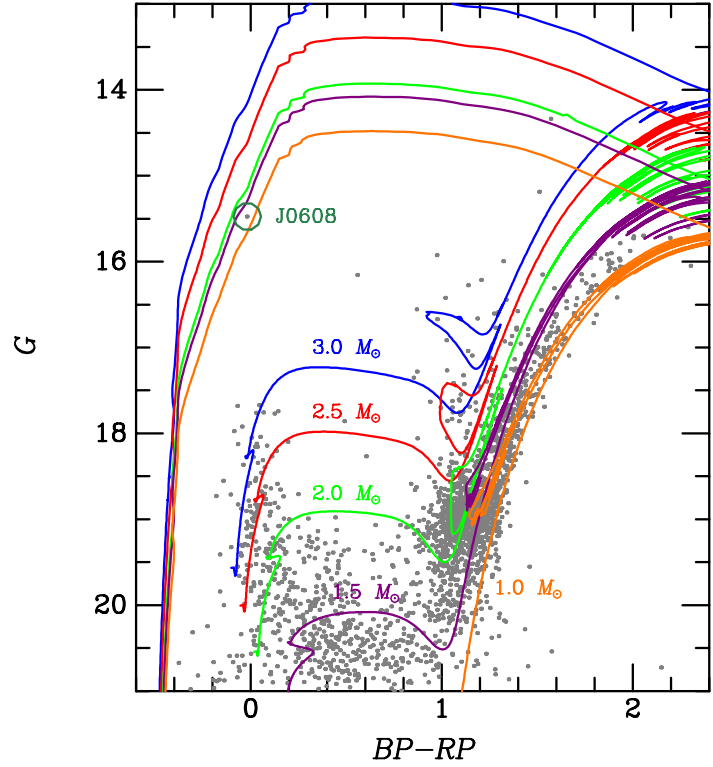}
\caption{
Color-magnitude diagram in the \Gaia\/ bandpasses of LMC stars within $10'$ of
J0608 ({\it gray filled circles}), selected from \Gaia\/ DR2 data on the basis of
proper motions as described in the text. Superposed ({\it colored lines}) are
theoretical stellar evolutionary tracks from the MIST website, for initial
masses of 1.0 to $3.0\,M_\odot$, and adjusted to the extinction, metallicity,
and distance of the LMC, as explained in the text. Also shown is the position of
J0608 itself ({\it large green circle}), but it would need to be adjusted to the left
and upwards for emission lines and circumstellar dust for comparison with the
tracks. There are stars as massive as $\sim\!3\,M_\odot$ in this field, and
the position of J0608 is consistent with it being descended from a fairly
massive progenitor star, and now lying on a post-AGB track at high luminosity.
\label{fig:cmd}
}
\end{figure}

To investigate the host LMC population near J0608, we selected all stars in
\Gaia\/ DR2 lying within an angular radius of $10'$ of the object, and with $\mu$
(in mas\,yr$^{-1}$) falling in the ranges $0.945<\mu_\alpha<2.445$ and
$0.233<\mu_\delta<1.733$. These criteria provide a nearly pure sample of stars
in the halo of the LMC at this location. In Figure~\ref{fig:cmd} we plot the
color-magnitude diagram (CMD) for these stars in the \Gaia\/ ($G_{BP}-G_{RP}$)
color index and $G$ apparent magnitude. The figure shows a well-populated giant
branch with a pronounced red-giant clump, as well as main-sequence stars
extending up to a color index of about zero.

To interpret this population, we have overlain the CMD data with theoretical
stellar evolutionary tracks obtained from the MESA Isochrones and Stellar Tracks
(MIST, version 1.2; Dotter 2016; Choi et al.\ 2016)
website\footnote{\url{http://waps.cfa.harvard.edu/MIST/}} and its web
interpolator. The MIST database includes calculations of magnitudes in the three
\Gaia\/ bandpasses. We adopted MIST's default stellar rotation of $v/v_{\rm
crit} = 0.4$, and obtained tracks for stars with initial masses of 1.0 to
$3.0\,M_\odot$, in steps of $0.5\,M_\odot$. We adjusted the tracks to an LMC
distance modulus of $(m-M)_0=18.48$~mag. After some experimentation, we found
that we could fit the population reasonably well using a metallicity of
$\rm[Fe/H]=-0.75$ (appropriate for this location in the LMC halo) and a
foreground extinction of $A_V=0.3$~mag (consistent with the reddening discussed
in \S1). 

Figure~\ref{fig:cmd} indicates that there are stars as massive as
$\sim\!3\,M_\odot$ at the site of J0608. We have also plotted, at the upper left
of the figure, the location of J0608 itself as measured in \Gaia\/ DR2. For
comparison with the theoretical tracks of normal stars, we would need to correct for the
emission lines as well as for the substantial circumstellar extinction. These
corrections---which we have not attempted in this initial discovery
paper---would move J0608 to the left and upward in the figure. Its position
would likely agree with the post-asymptotic-branch (post-AGB) tracks of the more
massive stars in this population.

However, the high carbon abundance and likely hydrogen deficiency of J0608 strongly suggest that it is not in a normal hydrogen-burning post-AGB evolutionary state. Evolutionary scenarios for late [WC] objects like \CPD\ have been discussed by many authors, including M\'endez (1991), Cohen et al.\ (1999), Crowther (2008), G{\'o}rny (2008), Clayton et al.\ (2011), Guerrero et al.\ (2018), and references therein. These scenarios usually involve a helium thermal pulse during post-AGB evolution. If the pulse occurs as late as the beginning of the descent of the white-dwarf cooling sequence, the star will retrace its post-AGB evolutionary track---such as those shown at the top of Figure~\ref{fig:cmd}---back to the red-giant regime. Thus J0608 could now be returning from such an event---or possibly it is still on its way to becoming a ``born-again'' red giant. Alternative scenarios involving binary interactions have also been discussed (e.g., De~Marco 2008; Clayton et al.\ 2011; Manick et al.\ 2015). We return to the question of binarity of J0608 in \S8. 

In summary, the radial-velocity, proper-motion, and parallax
measurements make it appear likely that J0608 belongs to the LMC, and that it is descended from a
star of initial mass possibly as high as $\sim\!3\,M_\odot$. The post-AGB evolution of
stars this massive is very rapid, and J0608 is in a transient evolutionary
stage of high, although not unprecedented, luminosity (cf.\ Sch\"onberner 1981).

 For the sake of completeness, we note that van Aarlie et al.\ (2011) have described two LMC stars that they suggest are of late [WC] spectral type; one is the well-studied HV~2671.  Although their spectra do contain \CII{} emission lines, they otherwise do not strongly resemble J0608, \CPD{}, or other objects in the WG74 class. As in the case of V348~Sgr discussed in \S3, we believe these may be more appropriately described as alternative objects such as R~CrB stars, in agreement with the classification already given for HV~2671 by multiple authors, e.g., Alcock et al.\ (2001), Soszy\'nki et al.\ (2009), and Clayton et al.\ (2011). However, we do note that the second of the Aarlie et al.\ (2011) objects, J055825.96$-$694425.8, has been termed ``likely a hot proto-PN or PN" by Hrivnak et al.\ (2015), although the spectrum presented by van Aarlie et al.\ (2011) appears to show prominent Mg~{\it b} absorption, again in contrast to the WG74 objects.

\section{Relation to Planetary Nebulae}

As all known Galactic late [WC] stars are surrounded by resolved PNe, one might wonder about the presence of a nebula associated with J0608, which might or might not be detectable at the LMC distance. J0608 is unresolved in SDSS $r$ and narrowband H$\alpha$ filters at the $\sim\!1\arcsec$ level on CCD direct images kindly obtained at our request by M.~Phillips (private communication) on multiple nights at both the du~Pont and Swope telescopes of the Las Campanas Observatory.  Any nebula at this location must be reasonably inconspicuous, because this region has long been well surveyed for PNe (e.g., Morgan 1994 and references therein), and multiple LMC-member PNe (as well as clusters) are indeed known even at this relatively large offset.  The PN MGPN~LMC86 (Morgan \& Good 1992), for example, is located just $18\arcmin$ distant, and is considerably fainter than J0608 (Leisy et al.\ 1997), perhaps leading one to wonder how the emission spectrum of J0608 has escaped previous attention, especially as there is no evidence for substantial photometric variability (see \S7) that might have pushed it beyond the limit of detectability in objective-prism surveys. However, the photographic emulsion/filter combination used in the MG survey (IIIa-J/GG455) by misfortune excludes the prominent emission lines in J0608.

Through the kind assistance of N. Hambly, we have examined a digitized version of the AAO/UKST SuperCOSMOS H$\alpha$ Survey film (Parker et al.\ 2005) which includes J0608. No nebulosity is apparent on angular scales ranging from a few arcseconds to several arcminutes.

The J0608 spectrum does show the clear presence of \fOII{}, \fNII{}, and \fSII{} emission, quite possibly originating in a surrounding but inconspicuous nebula. The intensity ratios of the \fOII{} and \fSII{} doublets indicate electron densities in the region where these lines are formed to be in the low-density limits of both doublets, i.e., $n<10^3$~cm$^{-3}$. This suggests the presence of a low density circumstellar nebular gas that is separate from the much higher density wind. The absence of any \fOIII{} $\lambda\lambda$4959, 5007 emission in J0608 does not alter this inference---it is also absent in other members of the class known to host easily resolved, low-excitation planetary nebula (e.g., De Marco, Barlow, \& Storey 1997; Henry et al.\ 2010).  We have examined the two-dimensional MagE spectrum at the \fOII{} lines to search for spatial extension, and find none, but the modest seeing at the time of this exposure makes this result inconclusive. Finally, the observed intense \WISE\/ IR flux for J0608 indicates that it shares the same marked IR excess common to late [WC] stars, which are known PN central stars (e.g., Zijlstra 2001; G\'{o}rny et al.\ 2001).

The lack of a detectable nebula in J0608, while disappointing, may not be surprising. Only a few LMC PNe have angular extent $>$$1\arcsec$, and the Galactic [WC11] nebulae, when extrapolated to the LMC distance, would be quite small in angular extent (e.g., De Marco et al.\ 2002; Chesneau et al.\ 2006).

\section{Photometric Variability}
\begin{figure}
\centering
\includegraphics[scale=.4]{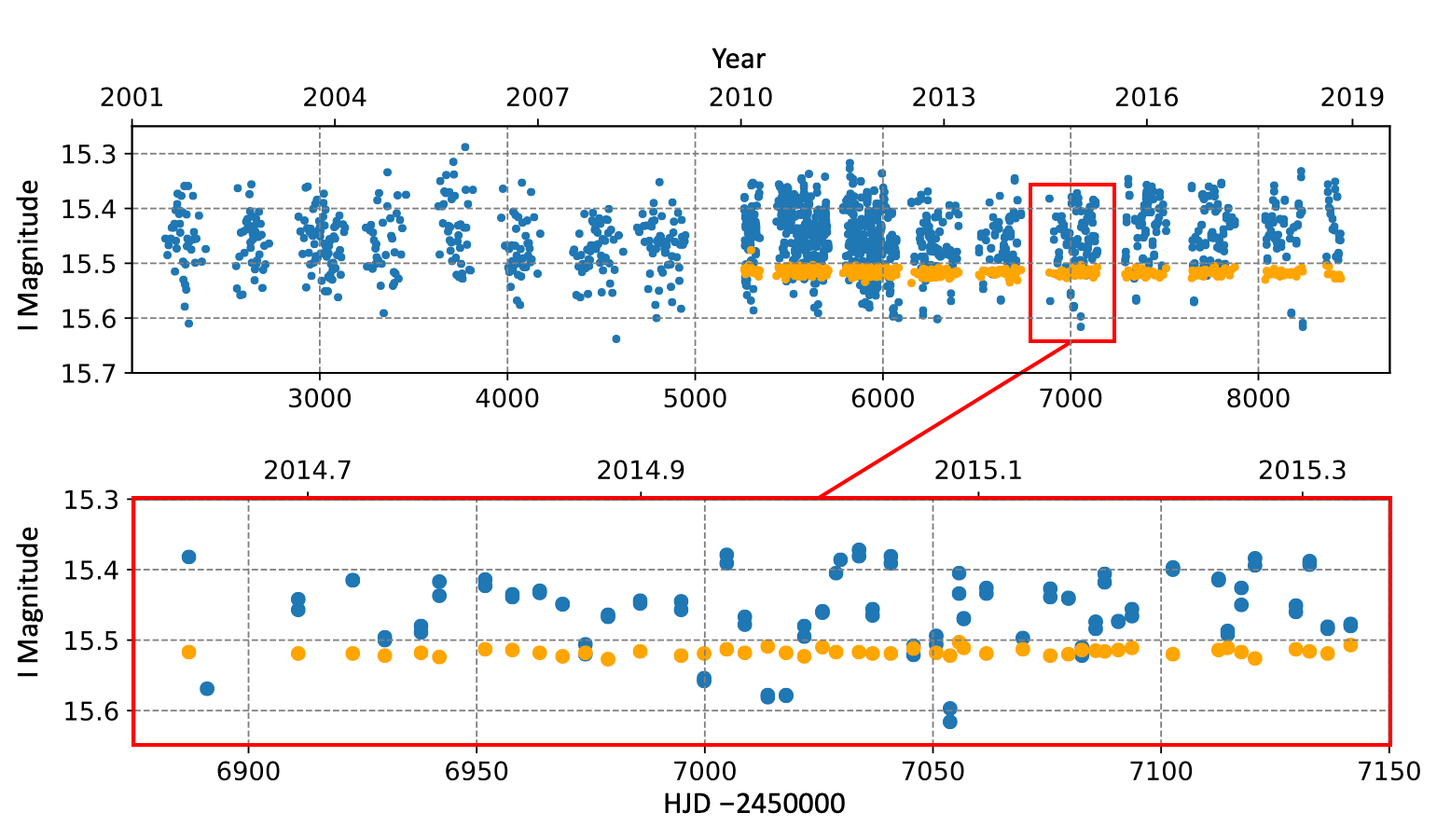}
\caption{OGLE $I$-band photometry of J0608 (blue dots) and a nearby comparison star of similar magnitude (orange dots) makes it clear that J0608 has significant, chaotic flickering, on a timescale of days, presumably due to a variable wind. The enlargement (lower panel) is an arbitrary section of data intended to show more detail. Time series analysis uncovers no significant periodic behavior.\label{fig:f6}}
\end{figure}

Due to its proximity to the LMC, J0608 has extensive broad-band photometry in the OGLE III and IV surveys (Udalski et al.\ 2015). The majority of the data are in the $I$-band, with some additional $V$ observations. The $I$-band data consist of observations on 946 nights in the 2001-2018 interval. Typical observations are shown in Figure~6. Over the long term, the object averages $I=15.45$, but irregular variations of up to $\sim$0.15 mag on timescales of days and weeks, well beyond the uncertainty in the measurements, are quite evident. The data were examined for periodicities in the 0--48 d$^{-1}$ range, but no significant periods were found. There are also ASAS-SN data (Shappee et al. 2014; Kochanek et al. 2017) on this star, exhibiting similar behavior.

Irregular variability of PN nuclei on a timescale of hours to days, as we observe here, has been reported for multiple objects (e.g., Bond \& Ciardullo 1991; Handler et al.\ 2013) and is typically ascribed to a variable stellar wind. The existence of a wind in J0608 is also supported by the dust implied by the bright mid-IR flux noted above, as well as the prominent P~Cyg  line profiles.

Historical photometry of J0608 is available from the Harvard plate collection via the DASCH archive (Grindlay et al.\ 2012; Tang et al.\ 2013). There are several dozen points spanning a century, with reasonably densely sampled data from 1925 to 1950. The object averages m$_{pg}\simeq15.3$, with no obvious variations beyond the $\pm$0.2~mag level. (With this statement we ignore a single highly discrepant, bright point, where inspection of the underlying image strongly suggests a spurious observation.) Although in limited, irregular sampling there is always the possibility of missing a sporadic event, this behavior is likely quite different from that of V348~Sgr, to which J0608 bears some spectral resemblance. The former star undergoes multiple dramatic ($\sim$6 mag) variation events (Heck et al.\ 1985), a signature of the R CrB class.

\section{The Photospheric Spectrum}

\begin{figure}
\centering
\includegraphics[scale=.7]{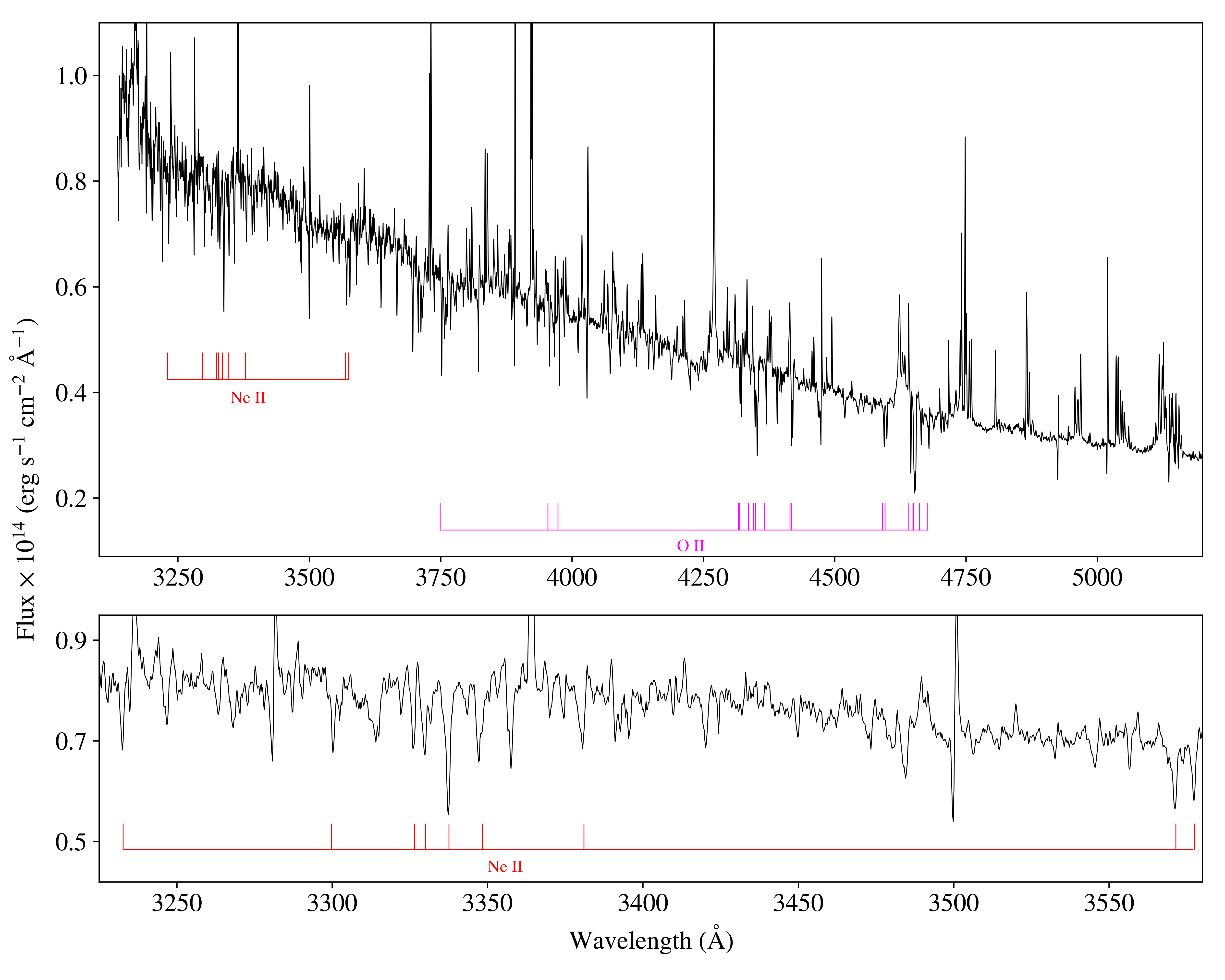}
\caption{{\it Upper panel:} An enlargement of the UV/blue spectrum of J0608 to illustrate the numerous stellar absorption lines. \NeII{} and \OII{} lines are marked. Most of the remaining absorption lines are \HeI{}. These data are identical to those of Fig.~1. {\it Lower panel:} A section of the spectrum encompassing the \NeII{} lines, yet further enlarged to show the profiles.\label{fig:zoomjspec}}
\end{figure}

Although the spectrum of J0608 is dominated by the intense emission of \CII{} and \HeI{}, which together with the associated P~Cyg profiles are normally attributed to a wind, there are also spectral clues to the underlying star. Shortward of $\sim$5200 \AA, absorption lines become noticeably more frequent, and the P~Cyg profiles are largely absent. We have identified these numerous absorption lines in the blue as \HeI{} and, remarkably, multiple \NeII{} and \OII{} lines (Figure~7), which suggest the presence of a hot subdwarf. The spectrum (and blue continuum) well match the known characteristics of hot subdwarf stars, for example Heber (2001). The absorption lines are perhaps narrower than one might expect for a typical subdwarf, possibly indicative of the high luminosity inferred in \S5, and lower surface gravity. \NeII{} absorption has been known in hot stars for almost a century (Menzel \& Marshall 1933), and these lines are another signature of hot subdwarf stars (Schindewolf et al.\ 2018). The lack of absorption longward of 5200~\AA\ is readily understandable. Such absorption multiplets exist but have low $f$-values, because they consist of absorption to $S$ states, which have lower energy than the $P$ and $D$ states, and hence longer wavelengths. Our (admittedly crude) estimates of $T_{\rm eff}$ in \S4 are also compatible with the hot-subdwarf hypothesis. In this scenario, the observed velocity differences between emission and absorption are due to the geometry of the star and its wind.

The colors of J0608 agree well with the large sample of hot subdwarf stars with \Gaia\/ parallaxes studied by Geier et al.\ (2019). The majority, although not quite all, of the stars in their sample are less luminous than we would infer for J0608 in the LMC\null. However, as discussed in \S5, there are hot post-AGB stars that briefly pass through the required luminosity, and examples are certainly known in the field (Sch\"onberner 1981).

Although the simultaneous presence of spectral lines of such disparate temperatures as seen in J0608 may seem odd, there is certainly precedent in the late [WC] stars.  IRAS 21282+5050, for example, is found to have characteristics of both a [WC11] and also O7(f) (Cohen \& Jones 1987, although the spectrum lacks \CII{} $\lambda$4267), and later classified simply as O9--O9.5 (Crowther et al.\ 1998).

An increasing fraction of PN central stars have been found to be binaries in recent years, but little is known about binarity in [WC] stars. Manick et al.\ (2015) report a spectroscopic binary in a [WC4] star, but failed to find evidence for binaries in multiple late-type [WC] stars similar to J0608. Miszalski et al.\ (2011) present more indirect evidence for the possibility that \CPD{} is a long-period (years) binary. We have compared our 2017 and 2019 MagE spectra of J0608 in a search for radial-velocity variability. The emission- and absorption-line mean velocities in the 2017 data have been quoted in \S3. The analogous values from the 2019 spectra are $+232.1 \pm 1.5$ \kms{} and $+211.6 \pm 3.2$ \kms{}, respectively. Thus in the time intervening between the two spectra, the velocity of the emission system has decreased by $15.0 \pm 2.1$ \kms{}, and the absorption system has decreased by a similar though not statistically significant amount, $6.0 \pm 5.4$ \kms{}. Therefore there is suggestive evidence for radial-velocity variability. Further radial-velocity monitoring is obviously desirable, and until variations are firmly detected, we cannot determine unequivocally whether J0608 is a single star or binary system.


\section{The Remarkable Excitation of \CII}

Previous comments on the interesting physics of the intense \CII{} emission in late [WC] stars have been made by several authors, e.g., Dahari \& Osterbrock (1984), Leuenhagen \& Hamann (1994), and De Marco, Storey, \& Barlow (1997). Although most authors attribute the lines to recombination (e.g., Kaler 1981), multiple other mechanisms are possible (Clegg 1989). Dahari \& Osterbrock (1984), for example, briefly discuss resonance fluorescence pumping, which we further expound upon below. 
   
The prominent emission lines in [WC] stars, which are formed in a stellar wind, differ from those of the emission spectra of gaseous nebulae, in that the strongest CNO lines are from high-excitation levels that are unlikely to be populated by collisional excitation. This is probably due to higher densities in the [WC] line-forming region compared with those of most nebulae, so the forbidden lines are collisionally de-excited. In addition, and in contrast to nearly all other emission-line objects, many of the emission lines observed in [WC] spectra originate from autoionizing states of \CII{}. A high abundance of carbon enables the normally weaker dielectronic recombination (DR) lines to be detectable in [WC] stars such as J0608, and an additional factor may be that excitation of the autoionizing levels may occur as a result of processes that are favored in stellar winds, such as radiative excitation, because the stellar radiation field is not diluted, as it is in nebulae. 
   
Among the prominent emission features in the J0608 spectrum that do not belong to H, \ion{He}{1}, or \CII{} are \OI{}  $\lambda7773$ and \OI{} $\lambda8446$, whose emission components have comparable strength and have been shown to be density diagnostics when they have comparable intensities. The upper level of $\lambda8446$, a triplet, is highly populated by strong resonance fluorescence from Ly$\beta$ scattering among excited \OI{} triplet states (Bhatia \& Kastner 1995). However,  \OI{}  $\lambda7773$ is a quintet, and with its high excitation potential the only way it can be excited at levels comparable to the 3p~$^3$P upper level of $\lambda8446$ is by collisional transfer from that level to the 3p~$^5$P upper level of  $\lambda7773$, which requires densities $n_e\ge10^{11}$~cm$^{-3}$ (Kastner \& Bhatia 1995). In spite of the fact that the strong absorption component of $\lambda7773$  is comparable to its emission component and thus causes the net line flux to be nil, the fact that the line is optically thick at its high excitation potential indicates that it is significantly coupled to the 3p~$^3$P level by high density. The primary line-emitting region of the J0608 wind may therefore be assumed to have densities in the regime of $n_e\ge10^{11}$~cm$^{-3}$, in accordance with density values that have been inferred from studies of other late [WC] stars (De~Marco \& Crowther 1998). 
   
An additional argument related to the density of the emitting region may be inferred from Balmer line intensities, inasmuch as recombination-line intensities for ionized gas at nebular densities are computed to have $I({\rm H}\alpha)/I({\rm H}\beta)\simeq2.7$ over a wide range of temperatures and densities (Storey \& Hummer 1995). It is seen from Table~1 that relative intensities are instead observed to be  $I({\rm H}\alpha)/I({\rm H}\beta)\approx2$ in both J0608 spectra, which are below the recombination value and therefore could be an indication of high densities, $n_e>10^{11}$~cm$^{-3}$ (Drake \& Ulrich 1980). However, it should be cautioned that a significant fraction of the H$\beta$ emission feature may be due to \CII{} $\lambda4862.6$, which could have an intensity comparable to that of H$\beta$, based on the intensity of its companion multiplet member \CII{} $\lambda4867$. Thus, information about densities inferred from the Balmer line ratios is uncertain. 
   
Detailed study of the spectrum of J0608 reveals that it is particularly complex. Like other [WC] stars, it shows an emission spectrum with prominent \HeI{} and \CII{} lines, many of which have P~Cyg profiles with absorption components that indicate significant optical depth in the lines. The Balmer series is present and stronger than the \HeI{} lines, but does not show appreciable absorption components as do the \HeI{} lines. The emission features are superposed on a quite blue continuum at visible wavelengths, and many of the \CII{} emission lines can be identified as originating from quartet states of high excitation potential, with $\chi\ge18$~eV. 
   
The presence of observed quartet multiplets in emission is particularly interesting in that quartet levels of \CII{} cannot be populated by either normal two-body radiative or DR of \CIII{} in its ground state because capture of a spin-1/2 free electron by the \CIII{}  2s$^2$~$^1$S$_{\rm 0}$ singlet ground state can produce only \CII{} doublet states.  However, as noted by Dahari \& Osterbrock (1984), DR can populate \CII{} quartet states via the first excited (metastable) triplet level of \CIII{}, 2s2p~$^3$P$^{\rm o}$, following its excitation by collisional or photo-excitation from the $^1$S$_{\rm 0}$ ground state. DR is a resonant radiation-less process where the kinetic energy of the free electron must be the same as the potential energy of the doubly excited state that it occupies. This strict constraint on the free electron energy causes DR rates to be very sensitive to the electron temperature $T_e$. 
   
Returning to the presence in the J0608 spectrum of \CII{} lines from quartet levels that can be populated by DR, the high densities and the possible excitation of doublet levels by resonance fluorescence, both from the continuum and from scattering of \HeI{} resonance lines, raises the possibility that the dominant process populating \CII{} autoionizing quartet levels may not be solely DR\null. At the densities in the line-emitting region, collisional transfer between bound doublet and quartet levels having similar excitation potentials should be strong, similar to the collisional process that couples \OI\ $\lambda7773$ to $\lambda8446$. 
   
Since collisional exchange between bound levels of \CII{} is likely to be strong, the best means of determining the significance of DR in populating quartet states is observation of those emission lines whose upper levels are above-the-continuum autoionizing states of \CII{}, which are generally considered to be excited by DR\null. Fortunately, there are a number of such transitions in the visible region that are detected in J0608. These include multiplets $\lambda\lambda4010-4021$ ($^4$S-–$^4$P$^{\rm o}$), $\lambda\lambda7046-7064$ ($^4$S–-$^4$P$^{\rm o}$), 
$\lambda\lambda5479-5490$ ($^4$D$^{\rm o}$–-$^4$D),\
$\lambda\lambda5250-5260$ ($^4$F$^{\rm o}$-–$^4$D), and 
$\lambda\lambda4313-4326$ ($^4$P-–$^4$P$^{\rm o}$) 
among the quartet states, and multiplets 
$\lambda\lambda6095-6103$ ($^2$P-–$^2$D$^{\rm o}$), 
$\lambda\lambda4960-4965$ ($^2$P-–$^2$P$^{\rm o}$), and 
$\lambda\lambda8794-8800$ ($^2$D-–$^2$F$^{\rm o}$) among the doublet states. 
   
DR coefficients have been published for only some of the above multiplets, so a comparison between observed and predicted relative intensities of \CII{} multiplets that are likely due to recombination is confined to a limited sample of lines. The intensities of relevant \CII{} multiplets are tabulated in Table~2, together with values predicted for radiative recombination (RR) for the doublets, and DR for the quartets at  $T_e=10^4$~K.  The DR values calculated for the quartets include the excitation of \CIII{} into its lowest excited level, 2p~$^3$P$^{\rm o}$, for that temperature.  The table also includes emission lines in the PN IC~418, where lower densities of  $n_e\sim10^4$~cm$^{-3}$ make it a useful comparison with the J0608 wind. No extinction correction has been applied for J0608 intensities, as the strong blue continuum and the location-based value of $E(B -− V )=0.09$ for its line of sight out of the Galaxy are both suggestive of small reddening. 
   
A comparison of the doublet intensities observed in J0608 with those predicted for radiative recombination shows disagreements by factors of up to three among the different lines, and in the sense that all of the observed lines are stronger than predicted relative to $\lambda4267$. This is indicative that recombination is not the only operative process populating the upper levels of C$^+$. Several of the autoionization multiplets that are detected in the spectrum originate on levels $\sim$2~eV above the \CII{} ionization limit, which requires that the emission region kinetic electron temperature must be of order $T_e\ge10^{3.6}$~K  (Storey \& Sochi 2013) if DR is the primary process populating the autoionizing levels. The fact that specific lines from highly excited doublet and quartet states have significant absorption components indicates large optical depth for those lines and is evidence of a large population of highly excited levels. Unusual for emission-line objects, this is difficult to achieve if only recombination, with its relatively small cross sections compared to collisional processes, is populating the levels. 

It is notable that the P~Cyg \CII{} profiles in J0608 are present in the quartet but not the doublet states. Examination of Table~2 shows the observed intensities of the J0608 quartet transitions, relative to the doublets, to be significantly greater than predicted from DR\null. Some of the quartet multiplets show significant absorption, which is rarely observed and indicative of appreciable optical depth in these lines with their high excitation potentials. These facts indicate the quartet states are populated by a process other than solely DR, and, given their excitation potentials of $\ge18$~eV, it is unlikely to be collisional excitation from the ground state. The relatively high density of the wind line-emitting region indicated by the \OI{} $\lambda7773$ strength relative to $\lambda8446$,  suggests that collisional transfer is occurring from doublet levels to quartet states of similar excitation potential. 
   
With one exception, \CII{} multiplet 24, the line intensities for the PN~IC 418 (Table~2) show quite good agreement with values predicted by radiative and dielectronic recombination. In contrast, for J0608 the  doublet states are populated by a process in addition to radiative recombination. 
  
An alternative excitation process to consider is radiative fluorescence from scattering of H and He resonance lines, especially since this is known to be an important process in PNe (Bowen 1934), active galactic nuclei (Weymann \& Williams 1969), novae (Strittmatter et al.\ 1977), and compact X-ray sources (Margon \& Cohen 1978). Coincidence in the wavelengths of resonance lines belonging to different ions produces a strong coupling between excitation of certain ions. Any fluorescence process between different ions almost certainly is initiated by line or continuum absorption from the ground state, which therefore strongly favors population of the \CII{} doublet levels. However, in searching for wavelength overlaps between \HeI{} and \CII{} resonance lines we find no clear coincidences likely to produce significant transfer of line radiation between those two ions. 
   
Excitation of high levels, both bound and autoionizing, can occur from the absorption of UV continuum radiation such that absorption of continuum radiation by resonance lines is a plausible source of the J0608 emission line spectrum. Such absorption enables the direct excitation of autoionizing states of \CII{}, inasmuch as radiative absorption from the ground state via two-electron excitation is allowed with $f$-values approaching $f=0.01$ (van Hoof 2018; Kramida et al.\ 2018). Strong continuum radiation in the 500--600 \AA\ region can pump \CII{} autoionizing doublet states directly, with collisional transfer to quartet states possible at the high wind densities before autoionization or stabilizing radiative decay.  
  
In summary, without detailed calculations it is difficult to make definite statements about the processes that produce the J0608 emission spectrum. Certainly, H is depleted with respect to He, and C is enhanced. The extent of carbon enhancement depends on the degree to which radiative absorption processes dominate recombination in populating excited doublet levels, which in turn collisionally populate the quartet levels. Future abundance determinations from strong \CII{} emission spectra originating in stellar winds should take these processes into account and not rely solely on recombination as the primary excitation mechanism (e.g., Pe\~na et al.\ 1997, Aleman et al.\ 2019). 
   
Finally, we call attention to the presence of absorption lines in the J0608 spectrum that are not associated with accompanying emission. These lines occur predominantly in the spectrum shortward of 5200 \AA, i.e., in the region where the continuum becomes increasingly strong into the blue. There are precedents for such absorption in late [WC] spectra that originate in the [WC] photosphere and are not fully obscured by the stellar wind (Crowther et al.\ 1998). The most definite of these features in other [WC] stars occur at wavelengths of 6146.66, 5596.95, 5806.14 and 5816.79; 4652.1 and 4654.2; and 8029.2 \AA. We identify the $\lambda\lambda5806, 5817$ absorption with the \CIV{} $\lambda\lambda5801.35, 5811.97$ doublet, and others as \OIII{} λ5592.3 and the \CIII{} λλ4647.4, 4650.2, 4651.5 complex. These have been identified previously in a late [WC] star (Cohen \& Jones 1987). Additionally, as noted in \S8, there are a number of absorption features that are identified with \OII{} and \NeII{}, as listed in Table~1. 
   
The structure of the J0608 emitting region is clearly complex.  Both collisional transfer between states and fluorescence are likely to be responsible for the strength of \CII{} emission, and a high density is indicated for the wind, in agreement with the presence of \OI{} $\lambda7773$, $\lambda8446$ (well-known to be due to resonance fluorescence), requiring $n>10^{11}$~cm$^{-3}$. On the other hand, the prominent forbidden emission of O, N, and S requires low density, so an undetected PN may well explain the presence of these latter lines.  A fuller understanding of the physical conditions that produce the spectra of [WC] stars awaits detailed models that incorporate the processes that populate the different levels contributing to the emission.

\section{Conclusions}

J0608 is a new member of the rare [WC11] class, and likely the first such extragalactic member, thus with an accurately determined luminosity due to LMC membership.  It is probably the central star of a PN, as are all other known [WC 11] stars, although no nebulosity is currently detected. As the prototype [WC11] star, \CPD{}, at an (approximately determined) distance of $\sim$1.4~kpc (De Marco, Barlow, \& Storey 1997), has a prominent PN extending to $\sim7\arcsec$ from the central star (Chesneau et al.\ 2006), the lack of a detection in J0608 at the LMC distance is probably not concerning.

We present arguments that the complex \CII{} emission spectrum of J0608 and others in its class is formed within the high-density region of a wind, as indicated by the presence of a strong \OI{} $\lambda7773$ P~Cyg feature.  A significant number of emission lines originate from autoionizing levels of \CII{} and its quartet levels. Their relative intensities differ markedly from those calculated for dielectronic recombination. Collisional transfer between levels of similar excitation potential is important in populating states of different multiplicity, and fluorescence from the stellar continuum is suggested as a significant process that may account for the appreciable intensities of lines detected from \CII{} quartet and autoionizing levels. The assumption of radiative and dielectroniic recombination alone as the cause of the \CII{} emission for this class of star may well overestimate the C abundance.

Observational selection clearly favors discovery of these stars in the PN phase, but J0608 lacks at least an easily visible nebulae, and also perhaps the most prominent spectral signature, strong \fOIII{} emission. Nor is there large-amplitude photometric variability to draw attention to the object. Finally, narrow-band photometric surveys tuned to the normally strongest WC lines (Massey et al.\ 2017 and references therein) will fail to discover these objects. We conclude that the number of late-type [WC] stars, both Galactic and in the Local Group, could be substantially larger than presently known.

\startlongtable
\begin{deluxetable*}{ccc|cccc|cccc}
\tablecaption{Identification and Measurements of J0608 Spectral Lines}
\tablecolumns{11}
\tablenum{1}
\tablehead{
\multicolumn{3}{c|}{ } & \multicolumn{4}{c|}{2017 December 30} & \multicolumn{4}{c}{2019 May 3} \\
\colhead{ID} & \colhead{Trans.} & \multicolumn{1}{c|}{Type} & \colhead{$\lambda_{\rm observed}$} & \colhead{Rel. Flux} & \colhead{EW} & \multicolumn{1}{c|}{RV\tablenotemark{c}} & \colhead{$\lambda_{\rm observed}$} & \colhead{Rel. Flux} & \colhead{EW} & \colhead{RV\tablenotemark{c}} \\
 & & & \colhead{(\AA)} & (Rel. to H$\beta$)\tablenotemark{a} & \colhead{(\AA)} & \colhead{(km s$^{-1}$)} & \multicolumn{1}{|c}{(\AA)} & (Rel. to H$\beta$)\tablenotemark{b} & \colhead{(\AA)} & \colhead{(km s$^{-1}$)}} 
\startdata
\CII{} 3165.5 & $^2$D$^{\rm o}$--$^2$P & em &  &  &  &   & 3168.0 & 1.75 &  & 237 \\
\CII{} 3167.9 & $^2$D$^{\rm o}$--$^2$P & em &  &  &  &   & 3170.3 & 0.95 &  & 227 \\
\HeI{} 3187.7 & $^3$S--$^3$P$^{\rm o}$ & P Cyg & 3190.0 & 0.57 & --0.19 & 217 & 3189.9 & 0.47 & --0.18 & 207 \\
 &  & abs & 3201.5 &  & --0.15 &   & 3200.8 &  & --0.13 &   \\
\NeII{} 3230.1 & $^2$D--$^2$D$^{\rm o}$ & abs & 3232.8 &  & --0.24 & 251 & 3232.4 &  & --0.19 & 214 \\
 &  & em & 3236.4 & 0.34 &  &   & 3236.4 & 0.4 &  &   \\
 &  & abs & 3280.6 &  & --0.27 &   & 3280.6 &  & --0.22 &   \\
 &  & em & 3282.0 & 0.26 &  &   & 3281.8 & 0.36 &  &   \\
\NeII{} 3297.7 & $^4$P--$^4$D$^{\rm o}$ & abs & 3300.7 &  & --0.17 & 273 & 3300.3 &  & --0.12 & 237 \\
 &  & abs & 3302.2 &  & --0.02 &   & 3302.7 &  & --0.07 &   \\
\NeII{} 3323.7 & $^2$P--$^2$P$^{\rm o}$ & abs & 3326.5 &  & --0.22 & 253 & 3326.1 &  & --0.27 & 217 \\
\NeII{} 3327.2 & $^4$P--$^4$D$^{\rm o}$ & abs & 3330.0 &  & --0.14 & 253 & 3329.8 &  & --0.27 & 234 \\
\NeII{} 3334.8 & $^2$D$^{\rm o}$--$^4$P & abs & 3337.6 &  & --0.42 & 252 & 3337.3 &  & --0.53 & 225 \\
\NeII{} 3345.5 & $^2$D--$^2$P$^{\rm o}$ & abs & 3347.7 &  & --0.31 & 197 & 3347.4 &  & --0.32 & 170 \\
 &  & abs & 3357.8 &  & --0.28 &   & 3357.6 &  & --0.33 &   \\
 &  & em & 3364.2 & 1.13 &  &   & 3364.0 & 1.38 &  &   \\
 &  & abs & 3370.2 &  & --0.15 &   & 3370.3 &  & --0.13 &   \\
\NeII{} 3378.2 & $^2$P--$^2$P$^{\rm o}$ & abs & 3381.0 &  & --0.13 & 249 & 3380.4 &  & --0.3 & 195 \\
 &  & abs & 3391.8 &  & --0.14 &   & 3391.2 &  & --0.09 &   \\
 &  & abs & 3395.7 &  & --0.09 &   & 3395.6 &  & --0.1 &   \\
 &  & abs & 3420.4 &  & --0.22 &   & 3420.2 &  & --0.17 &   \\
\HeI{} 3447.3 & $^1$S--$^1$P$^{\rm o}$ & abs & 3449.8 &  & --0.08 & 218 & 3449.8 &  & --0.13 & 218 \\
 &  & abs & 3473.7 &  & --0.11 &   & 3472.8 &  & --0.14 &   \\
 &  & abs & 3484.5 &  & --0.23 &   & 3484.3 &  & --0.28 &   \\
\HeI{} 3498.6 & $^3$P$^{\rm o}$--$^3$D & P Cyg & 3500.5 & 0.38 & --0.18 & 163 & 3500.3 & 0.46 & --0.21 & 146 \\
 &  & abs & 3514.6 &  & --0.08 &   & 3514.7 &  & --0.03 &   \\
 &  & abs & 3545.6 &  & --0.11 &   & 3545.3 &  & --0.18 &   \\
\NeII{} 3568.5 & $^2$D--$^2$F$^{\rm o}$ & abs & 3571.6 &  & --0.16 & 261 & 3571.2 &  & --0.21 & 227 \\
\NeII{} 3574.6 & $^2$D--$^2$P$^{\rm o}$ & abs & 3577.6 &  & --0.19 & 252 & 3577.3 &  & --0.24 & 227 \\
\HeI{} 3587.3 & $^3$P$^{\rm o}$--$^3$D & abs & 3589.7 &  & --0.16 & 201 & 3589.6 &  & --0.16 & 192 \\
 &  & em & 3593.7 & 0.25 &  &   &  &  &  &   \\
 &  & em & 3604.6 & 0.17 &  &   &  &  &  &   \\
 &  & abs & 3613.7 &  & --0.04 &   & 3612.0 &  & --0.04 &   \\
\HeI{} 3613.6 & $^1$S--$^1$P$^{\rm o}$ & abs & 3615.9 &  & --0.05 & 191 & 3616.0 &  & --0.08 & 199 \\
 &  & abs & 3636.6 &  & --0.23 &   & 3636.5 &  & --0.28 &   \\
 &  & abs & 3667.1 &  & --0.3 &   & 3666.7 &  & --0.34 &   \\
 &  & abs & 3697.2 &  & --0.34 &   & 3696.9 &  & --0.47 &   \\
 &  & em & 3703.6 & 0.2 &  &   &  &  &  &   \\
 &  & abs & 3707.3 &  & --0.19 &   & 3707.4 &  & --0.16 &   \\
 &  & abs & 3712.9 &  & --0.1 &   & 3712.5 &  & --0.15 &   \\
\fOII{} 3726.0 & $^4$S$^{\rm o}$--$^2$D$^{\rm o}$ & em & 3728.9 & 1.06 &  & 234 & 3728.7 & 0.85 &  & 217 \\
\fOII{} 3728.8 & $^4$S$^{\rm o}$--$^2$D$^{\rm o}$ & em & 3731.6 & 1.7 &  & 225 & 3731.5 & 1.3 &  & 217 \\
 &  & abs & 3738.4 &  & --0.14 &   & 3737.8 &  & --0.12 &   \\
\OII{} 3749.5 & $^4$P--$^4$S$^{\rm o}$ & abs & 3752.0 &  & --0.31 & 200 & 3752.1 &  & --0.36 & 208 \\
 &  & abs & 3758.0 &  & --0.27 &   & 3757.5 &  & --0.2 &   \\
 &  & abs & 3769.5 &  & --0.18 &   & 3769.1 &  & --0.23 &   \\
\SiIII{} 3796.1 & $^3$P$^{\rm o}$--$^3$D & em & 3799.6 & 0.6 &  & 277 & 3799.2 & 0.48 &  & 245 \\
 &  & em & 3805.6 & 0.27 &  &   & 3805.4 & 0.22 &  &   \\
\SiIII{} 3806.6 & $^3$P$^{\rm o}$--$^3$D & em & 3809.8 & 0.38 &  & 252 & 3809.5 & 0.25 &  & 229 \\
\HeI{} 3818.9 & $^3$P$^{\rm o}$--$^3$D & abs & 3822.0 &  & --0.31 & 244 & 3822.0 &  & --0.27 & 244 \\
\CII{} 3831.7 & $^2$P$^{\rm o}$--$^2$D & em & 3834.8 & 0.64 &  & 243 & 3834.6 & 0.55 &  & 227 \\
\CII{} 3835.7 & $^2$P$^{\rm o}$--$^2$D & em & 3839.2 & 0.88 &  & 274  & 3839.0 & 0.68 &  & 258  \\
 \& H9 &  \& 2--9 &  &  &  &  &   &  &  &  &   \\
 &  & em & 3851.7 & 0.37 &  &   &  &  &  &   \\
 &  & em & 3858.6 & 0.24 &  &   & 3858.5 & 0.3 &  &   \\
 &  & em & 3880.4 & 0.72 &  &   &  &  &  &   \\
\HeI{} 3888.6 & $^3$S--$^3$P$^{\rm o}$ & P Cyg & 3891.3 & 1.87 & --0.3 & 208 & 3891.3 & 2.0 & --0.18 & 208 \\
\CII{} 3919.0 & $^2$P$^{\rm o}$--$^2$S & em & 3922.3 & 1.18 &  & 253 & 3922.0 & 0.91 &  & 230 \\
\CII{} 3920.7 & $^2$P$^{\rm o}$--$^2$S & em & 3924.0 & 2.2 &  & 253 & 3923.8 & 1.6 &  & 237 \\
 &  & em & 3927.1 & 0.25 &  &   & 3926.9 & 0.3 &  &   \\
 &  & em & 3932.1 & 0.24 &  &   & 3931.9 & 0.33 &  &   \\
\OII{} 3954.4 & $^2$P--$^2$P$^{\rm o}$ & abs & 3957.2 &  & --0.29 & 212 & 3957.2 &  & --0.37 & 212 \\
\HeI{} 3964.7 & $^1$S--$^1$P$^{\rm o}$ & P Cyg & 3967.5 & 0.15 & --0.14 & 212 & 3967.4 & 0.18 & --0.11 & 204 \\
H$\epsilon$ & 2-7 & em & 3973.4 & 0.14 &  &  251 & 3973.1 &  &  & 229  \\
\OII{} 3973.3 & $^2$P--$^2$P$^{\rm o}$ & abs & 3976.1 &  & --0.25 & 211 & 3976.2 &  & --0.33 & 219 \\
 &  & em & 3983.3 & 0.18 &  &   & 3983.3 & 0.13 &  &   \\
 &  & em & 3988.3 & 0.09 &  &   & 3988.3 & 0.1 &  &   \\
\HeI{} 4009.3 & $^1$P$^{\rm o}$--$^1$D & P Cyg & 4011.7 & \textless0.01 & --0.12 & 180 & 4012.3 & 0.08 & --0.17 & 225 \\
 &  & em & 4019.4 & 0.36 &  &   & 4019.0 & 0.2 &  &   \\
\HeI{} 4026.2 & $^3$P$^{\rm o}$--$^3$D & P Cyg & 4029.5 & 0.94 & --0.41 & 246 & 4029.4 & 1.08 & --0.37 & 238 \\
H$\delta$ & 2-6 & em & 4105.3 & 0.22 &  & 261  & 4105.0 & 0.25 &  & 239  \\
\HeI{} 4120.8 & $^3$P$^{\rm o}$--$^3$S & abs & 4123.4 &  & --0.11 & 189 & 4123.0 &  & --0.18 & 160 \\
 &  & em & 4126.9 & 0.11 &  &   & 4126.8 & 0.12 &  &   \\
 &  & em & 4131.9 & 0.27 &  &   & 4131.8 & 0.29 &  &   \\
 &  & em & 4135.1 & 0.46 &  &   & 4134.9 & 0.31 &  &   \\
\HeI{} 4143.8 & $^1$P$^{\rm o}$--$^1$D & abs & 4146.0 &  & --0.2 & 159 & 4146.0 &  & --0.23 & 159 \\
 &  & em & 4159.6 & 0.17 &  &   & 4159.4 & 0.12 &  &   \\
 &  & em & 4211.4 & 0.28 &  &   & 4211.1 & 0.2 &  &   \\
 &  & em & 4214.7 & 0.27 &  &   & 4214.5 & 0.31 &  &   \\
 &  & abs & 4224.6 &  & --0.19 &   &  &  &  &   \\
\CII{} 4267.3 & $^2$G--$^4$F$^{\rm o}$ & em & 4270.9 & 3.83 &  & 253 & 4270.7 & 3.43 &  & 239 \\
\CII{} 4292.3 & $^2$F$^{\rm o}$--$^2$G & em & 4295.8 & 0.28 &  & 245 & 4295.6 & 0.29 &  & 231 \\
\CII{} 4306.3 & $^2$P$^{\rm o}$--$^2$S & em & 4310.3 & 0.37 &  & 279 & 4310.1 &  &  & 265 \\
\OII{} 4317.1 & $^4$P--$^4$P$^{\rm o}$ & abs & 4320.1 &  & --0.13 & 209 & 4320.1 &  & --0.31 & 209 \\
\OII{} 4319.6 & $^4$P--$^4$P$^{\rm o}$ & abs & 4323.0 &  & --0.22 & 236 & 4322.9 &  & --0.36 & 229 \\
\CII{} 4329.8 & $^2$D--$^2$F$^{\rm o}$ & em & 4333.4 & 0.31 &  & 249 & 4333.2 & 0.44 &  & 236 \\
\OII{} 4336.9 & $^4$P--$^4$P$^{\rm o}$ & abs & 4339.8 &  & --0.05 & 201 & 4340.0 &  & --0.11 & 214 \\
H$\gamma$ & 2-5 & em & 4344.0 & 0.28 &  & 244  & 4343.8 & 0.29 &  & 231  \\
\OII{} 4345.6 & $^4$P--$^4$P$^{\rm o}$ & abs & 4348.6 &  & --0.21 & 207 & 4348.9 &  & --0.31 & 228 \\
\OII{} 4349.4 & $^4$P--$^4$P$^{\rm o}$ & abs & 4352.5 &  & --0.35 & 214 & 4352.6 &  & --0.34 & 221 \\
\OII{} 4366.9 & $^4$P--$^4$P$^{\rm o}$ & abs & 4369.8 &  & --0.36 & 199 & 4370.0 &  & --0.47 & 213 \\
\CII{} 4372.4 & $^4$P$^{\rm o}$--$^4$D & em & 4375.9 & 0.18 &  & 240 & 4375.6 & 0.18 &  & 220 \\
\CII{} 4375.0 & $^4$P$^{\rm o}$--$^4$D & em & 4378.3 & 0.17 &  & 226 & 4378.3 & 0.08 &  & 226 \\
\CII{} 4376.6 & $^4$P$^{\rm o}$--$^4$D & em & 4379.8 & 0.3 &  & 219 & 4379.7 & 0.18 &  & 213 \\
\HeI{} 4387.9 & $^1$P$^{\rm o}$--$^1$D & abs & 4390.6 &  & --0.34 & 185 & 4390.6 &  & --0.31 & 185 \\
\CII{} 4411.3 & $^2$D$^{\rm o}$--$^2$F & em & 4414.4 & 0.59 &  & 211 & 4414.2 & 0.63 &  & 197 \\
\OII{} 4414.9 & $^2$P--$^2$D$^{\rm o}$ & abs & 4418.4 &  & --0.33 & 238 & 4418.3 &  & --0.34 & 231 \\
\OII{} 4417.0 & $^2$P--$^2$D$^{\rm o}$ & abs & 4420.2 &  & --0.16 & 217 & 4420.4 &  & --0.25 & 231 \\
 &  & em & 4456.8 & 0.16 &  &   & 4457.0 & 0.1 &  &   \\
 &  & em & 4460.6 & 0.17 &  &   & 4460.5 & 0.21 &  &   \\
\HeI{} 4471.5 & $^3$P$^{\rm o}$--$^3$D & P Cyg & 4474.8 & 0.5 & --0.34 & 221 & 4474.5 & 0.53 & --0.13 & 201 \\
\CII{} 4491.2 & $^2$F$^{\rm o}$--$^2$G & em & 4495.0 & 0.32 &  & 254 & 4494.7 & 0.29 &  & 234 \\
 &  & abs & 4519.9 &  & --0.15 &   & 4519.6 &  & --0.17 &   \\
\OII{} 4591.0 & $^2$D--$^2$F$^{\rm o}$ & abs & 4594.2 &  & --0.19 & 209 & 4594.4 &  & --0.27 & 222 \\
\OII{} 4596.0 & $^2$D--$^2$F$^{\rm o}$ & abs & 4599.3 &  & --0.19 & 215 & 4599.4 &  & --0.29 & 222 \\
\CII{} 4619.2 & $^2$F$^{\rm o}$--$^2$G & em & 4623.4 & 0.92 &  & 273 & 4623.2 & 0.74 &  & 260 \\
\CII{} 4637.6 & $^2$P$^{\rm o}$--$^2$D & em & 4641.1 & 0.17 &  & 226 & 4640.9 & 0.29 &  & 214 \\
\OII{} 4641.8 & $^4$P--$^4$D$^{\rm o}$ & abs & 4645.2 &  & --0.52 & 220 & 4645.1 &  & --0.61 & 213 \\
\OII{} 4649.1 & $^2$F--$^4$D$^{\rm o}$ & abs & 4652.2 &  & --1.14 & 200 & 4652.6 &  & --0.85 & 226 \\
\OII{} 4650.8 & $^4$P--$^4$D$^{\rm o}$ & abs & 4654.2 &  & --0.57 & 219 & 4654.4 &  & --0.39 & 232 \\
\OII{} 4661.6 & $^4$P--$^4$D$^{\rm o}$ & abs & 4664.7 &  & --0.2 & 200 & 4664.9 &  & --0.22 & 212 \\
\OII{} 4676.2 & $^4$P--$^4$D$^{\rm o}$ & abs & 4679.2 &  & --0.2 & 193 & 4679.6 &  & --0.17 & 218 \\
 &  & em & 4700.2 & 0.2 &  &   & 4699.9 & 0.14 &  &   \\
\HeI{} 4713.2 & $^3$P$^{\rm o}$--$^3$S & P Cyg & 4716.5 & 0.26 & --0.07 & 210 & 4716.4 & 0.29 & --0.04 & 204 \\
\CII{} 4735.5 & $^2$P--$^2$P$^{\rm o}$ & em & 4739.5 & 0.34 &  & 253 & 4739.1 & 0.25 &  & 228 \\
\CII{} 4738.0 & $^2$P--$^2$P$^{\rm o}$ & em & 4741.9 & 0.67 &  & 247 & 4741.6 & 0.79 &  & 228 \\
\CII{} 4744.8 & $^2$P--$^2$P$^{\rm o}$ & em & 4748.7 & 1.11 &  & 247 & 4748.5 & 1.14 &  & 234 \\
\CII{} 4747.3 & $^2$P--$^2$P$^{\rm o}$ & em & 4751.1 & 0.35 &  & 240 & 4750.8 & 0.43 &  & 221 \\
 &  & em & 4756.7 & 0.29 &  &   & 4756.5 & 0.43 &  &   \\
 &  & em & 4760.4 & 0.36 &  &   & 4760.3 & 0.48 &  &   \\
\CII{} 4802.7 & $^2$F$^{\rm o}$--$^2$G & em & 4806.4 & 0.41 &  & 231 & 4806.2 & 0.41 &  & 219 \\
\CII{} 4862.6 & $^2$S-$^2$P$^0$ & em & 4865.6 & 1.0 &  & 185  & 4865.5 & 1.0 &  &  179 \\
 \& H$\beta$ &  \& 2-4 &  &  &  &  &   &  &  &  &   \\
\CII{} 4867.1 & $^2$S--$^2$P$^{\rm o}$ & em & 4870.8 & 0.23 &  & 228 & 4870.6 & 0.28 &  & 216 \\
\HeI{} 4921.9 & $^1$P$^{\rm o}$--$^1$D & P Cyg & 4925.5 & 0.14 & --0.42 & 219 & 4925.3 & 0.17 & --0.34 & 207 \\
 &  & em & 4944.9 & 0.09 &  &   & 4944.7 & 0.06 &  &   \\
\CII{} 4953.9 & $^2$P--$^2$P$^{\rm o}$ & em & 4957.7 & 0.16 &  & 230 & 4957.5 & 0.22 &  & 218 \\
\CII{} 4958.7 & $^2$P--$^2$P$^{\rm o}$ & em & 4963.2 & 0.23 &  & 272 & 4962.9 & 0.26 &  & 254 \\
\CII{} 4964.7 & $^2$P--$^2$P$^{\rm o}$ & em & 4968.7 & 0.44 &  & 242 & 4968.5 & 0.46 &  & 230 \\
\HeI{} 5015.7 & $^1$S--$^1$P$^{\rm o}$ & P Cyg & 5019.1 & 0.85 & --0.29 & 203 & 5019.0 & 0.88 & --0.17 & 197 \\
\CII{} 5032.1 & $^2$P$^{\rm o}$--$^2$D & em & 5036.2 & 0.57 &  & 244 & 5035.8 & 0.44 &  & 221 \\
\CII{} 5035.9 & $^2$P$^{\rm o}$--$^2$D & em & 5039.9 & 0.53 &  & 238 & 5039.6 & 0.48 &  & 220 \\
\CII{} 5040.7 & $^2$P$^{\rm o}$--$^2$D & em & 5044.9 & 0.25 &  & 250 & 5044.6 & 0.31 &  & 232 \\
\CII{} 5044.4 & $^4$D$^{\rm o}$--$^4$P & em & 5048.3 & 0.21 &  & 232 & 5048.3 & 0.19 &  & 232 \\
\HeI{} 5047.7 & $^1$P$^{\rm o}$--$^1$S & em & 5052.0 & 0.17 &  & 256 & 5052.0 & 0.16 &  & 256 \\
\CII{} 5114 & $^2$P$^{\rm o}$--$^2$D & em & 5118.1 & 0.56 &  & 241 & 5117.9 & 0.62 &  & 229 \\
\CII{} 5120.1 & $^2$P$^{\rm o}$--$^2$P & em & 5124.0 & 0.4 &  & 229 & 5123.5 & 0.16 &  & 199 \\
\CII{} 5121.8 & $^2$P$^{\rm o}$--$^2$P & em & 5126.1 & 0.47 &  & 252 & 5125.7 & 0.35 &  & 228 \\
\CII{} 5125.2 & $^2$P$^{\rm o}$--$^2$P & em & 5129.4 & 0.11 &  & 246 & 5129.1 & 0.12 &  & 228 \\
\CII{} 5132.9 & $^4$P$^{\rm o}$--$^4$P & abs & 5136.2 &  & --0.34 & 193 & 5136.2 &  & --0.28 & 193 \\
\CII{} 5133.3 & $^4$P$^{\rm o}$--$^4$P & em & 5138.0 & 0.27 &  & 275 & 5137.8 & 0.27 &  & 263 \\
\CII{} 5137.3 & $^4$P$^{\rm o}$--$^4$P & em & 5141.8 & 0.3 &  & 263 & 5141.4 & 0.12 &  & 239 \\
\CII{} 5139.2 & $^4$P$^{\rm o}$--$^4$P & em & 5143.5 & 0.35 &  & 251 & 5143.4 & 0.2 &  & 245 \\
\CII{} 5143.5 & $^4$P$^{\rm o}$--$^4$P & abs & 5146.7 &  & --0.26 & 187 & 5146.7 &  & --0.09 & 187 \\
\CII{} 5145.2 & $^4$P$^{\rm o}$--$^4$P & P Cyg & 5149.2 & 0.24 & --0.09 & 233 & 5149.1 & 0.22 & --0.09 & 227 \\
\CII{} 5151.1 & $^4$P$^{\rm o}$--$^4$P & P Cyg & 5154.9 & 0.21 & --0.24 & 221 & 5154.9 & 0.17 & --0.15 & 221 \\
\CII{} 5332.9 & $^2$P$^{\rm o}$--$^2$S & em & 5337.1 & 0.3 &  & 236 & 5336.6 & 0.13 &  & 208 \\
\CII{} 5334.8 & $^2$P$^{\rm o}$--$^2$S & em & 5338.9 & 0.36 &  & 231 & 5338.5 & 0.32 &  & 208 \\
\CII{} 5342.4 & $^2$F$^{\rm o}$--$^2$G & em & 5346.7 & 0.43 &  & 242 & 5346.4 & 0.4 &  & 225 \\
 &  & em & 5372.6 & 0.23 &  &   & 5372.3 & 0.2 &  &   \\
 &  & em & 5375.4 & 0.06 &  &   & 5374.6 & 0.13 &  &   \\
 &  & em & 5379.5 & 0.14 &  &   & 5378.9 & 0.06 &  &   \\
\CIV{} 5411.4 & $^2$G--$^2$H$^o$ & abs & 5416.8 &  & --0.03 & 299  & 5415.9 &  & --0.07 & 249 \\
 \& \HeII{} 5411.5 &  \& 4-7 &  &  &  &  &   &  &  &  &   \\
 &  & em & 5483.0 & 0.12 &  &   & 5482.7 & 0.09 &  &   \\
\OIII{} 5508.2 & $^1$D--$^1$D$^o$ & em & 5513.2 &  & --0.06 & 272 & 5512.6 &  & --0.05 & 240 \\
\CII{} 5535.4 & $^2$S--$^2$P$^{\rm o}$ & em & 5540.4 & 1.04 &  & 271 & 5539.5 & 0.29 &  & 222 \\
\CII{} 5537.6 & $^2$S--$^2$P$^{\rm o}$ & em & 5541.8 & 0.45 &  & 228 & 5541.8 & 0.34 &  & 228 \\
\OIII{} 5592.3 & $^1$P$^{\rm o}$--$^1$P & abs & 5597.0 &  & --0.27 & 252 & 5596.3 &  & --0.29 & 215 \\
\CII{} 5640.5 & $^4$P$^{\rm o}$--$^4$S & em & 5645.4 & 0.3 &  & 261 & 5645.2 & 0.22 &  & 250 \\
\CII{} 5648.1 & $^4$P$^{\rm o}$--$^4$S & P Cyg & 5652.1 & 0.23 & --0.06 & 213 & 5651.9 & 0.25 & --0.05 & 202 \\
\CII{} 5662.5 & $^4$P$^{\rm o}$--$^4$S & P Cyg & 5666.6 & 0.25 & --0.12 & 217 & 5666.5 & 0.18 & --0.08 & 212 \\
\CIII{} 5695.9 & $^1$P$^{\rm o}$--$^1$D & em & 5700.7 & 0.15 &  & 253 & 5701.3 & 0.05 &  & 284 \\
 &  & em & 5744.4 & 0.06 &  &   & 5744.2 & 0.07 &  &   \\
\CIV{} 5801.4 & $^2$S--$^2$P$^{\rm o}$ & abs & 5806.2 &  & --0.21 & 248 & 5805.5 &  & --0.24 & 212 \\
\CIV{} 5812.0 & $^2$S--$^2$P$^{\rm o}$ & abs & 5816.9 &  & --0.26 & 253 & 5816.2 &  & --0.15 & 217 \\
\CII{} 5818.3 & $^4$D--$^4$P$^{\rm o}$ & em & 5823.1 & 0.06 &  & 248 & 5822.4 & 0.06 &  & 211 \\
\CII{} 5823.2 & $^4$D--$^4$P$^{\rm o}$ & em & 5827.6 & 0.09 &  & 227 & 5827.4 & 0.08 &  & 216 \\
\CII{} 5827.9 & $^4$D--$^4$P$^{\rm o}$ & em & 5832.4 & 0.09 &  & 232 & 5832.2 & 0.1 &  & 221 \\
\CII{} 5836.4 & $^4$D--$^4$P$^{\rm o}$ & em & 5841.0 & 0.11 &  & 236 & 5840.6 & 0.14 &  & 216 \\
\CII{} 5843.6 & $^4$D--$^4$P$^{\rm o}$ & em & 5848.6 & 0.07 &  & 257 & 5848.4 & 0.15 &  & 246 \\
\CII{} 5856.1 & $^4$D--$^4$P$^{\rm o}$ & em & 5860.9 & 0.18 &  & 246 & 5860.5 & 0.21 &  & 225 \\
\HeI{} 5875.6 & $^3$P$^{\rm o}$--$^3$D & P Cyg & 5879.6 & 1.77 & --0.33 & 204 & 5879.4 & 1.77 & --0.13 & 194 \\
\CII{} 5889.8 & $^2$D--$^2$P$^{\rm o}$ & em & 5895.2 & 2.81 &  & 275 & 5894.8 & 2.71 &  & 255 \\
\CII{} 5907.2 & $^4$P$^{\rm o}$--$^4$S & em & 5911.9 & 0.07 &  & 239 & 5911.6 & 0.09 &  & 224 \\
\CII{} 5914.6 & $^4$P$^{\rm o}$--$^4$S & em & 5919.8 & 0.08 &  & 264 & 5919.3 & 0.11 &  & 238 \\
 &  & em & 6011.5 & 0.06 &  &   & 6011.2 & 0.11 &  &   \\
 &  & em & 6024.6 & 0.07 &  &   & 6024.3 & 0.08 &  &   \\
 &  & em & 6039.1 & 0.04 &  &   & 6038.6 & 0.04 &  &   \\
 &  & em & 6042.0 & 0.08 &  &   & 6041.9 &  &  &   \\
 &  & em & 6067.2 & 0.06 &  &   & 6067.0 & 0.08 &  &   \\
 &  & em & 6083.5 & 0.06 &  &   & 6083.2 & 0.09 &  &   \\
\CII{} 6095.3 & $^2$P--$^2$D$^{\rm o}$ & em & 6100.5 & 0.63 &  & 256 & 6100.0 & 0.34 &  & 231 \\
\CII{} 6098.5 & $^2$P--$^2$D$^{\rm o}$ & em & 6103.4 & 0.52 &  & 241 & 6103.1 & 0.57 &  & 226 \\
\CII{} 6102.6 & $^2$P--$^2$D$^{\rm o}$ & em & 6107.4 & 0.22 &  & 236 & 6107.4 & 0.15 &  & 236 \\
 &  & em & 6118.8 & 0.08 &  &   & 6118.7 & 0.09 &  &   \\
\CII{} 6151.5 & $^2$D--$^2$F$^{\rm o}$ & em & 6156.5 & 0.67 &  & 244 & 6156.2 & 0.59 &  & 229 \\
 &  & em & 6172.3 & 0.08 &  &   & 6171.9 & 0.16 &  &   \\
 &  & em & 6181.0 & 0.05 &  &   & 6180.8 & 0.08 &  &   \\
 &  & em & 6218.0 & 0.12 &  &   & 6218.0 & 0.12 &  &   \\
\CII{} 6251.1 & $^2$F$^{\rm o}$--$^2$G & em & 6256.2 & 0.15 &  & 245 & 6255.8 & 0.15 &  & 226 \\
\CII{} 6257.2 & $^2$P$^{\rm o}$--$^2$D & em & 6262.3 & 0.4 &  & 245 & 6261.8 & 0.22 &  & 221 \\
\CII{} 6259.6 & $^2$P$^{\rm o}$--$^2$D & em & 6264.5 & 0.51 &  & 235 & 6264.1 & 0.55 &  & 216 \\
 &  & em & 6284.9 & 0.06 &  &   & 6284.7 & 0.08 &  &   \\
\fOI{} 6300.3 & $^3$P--$^1$D & em & 6305.2 & 0.12 &  &  233 & 6305.0 & 0.1 &  & 224  \\
\SiII{} 6347.1 & $^2$S--$^2$P$^{\rm o}$ & em & 6351.6 & 0.18 &  & 213 & 6351.2 & 0.27 &  & 194 \\
\& \MgII{} 6346.8 & \& $^2$D--$^2$F$^{\rm o}$ &&&&&&&&&\\
 &  & em & 6362.4 & 0.03 &  &   & 6362.0 & 0.03 &  &   \\
 &  & em & 6365.1 & 0.01 &  &   &  &  &  &   \\
 &  & em & 6367.7 & 0.04 &  &   &  &  &  &   \\
\fOI{} 6363.8 & $^3$P--$^1$D & em & 6368.4 & 0.09 &  & 217 & 6368.3 & 0.15 &  & 212  \\
\SiII{} 6371.0 & $^2$S--$^2$P$^{\rm o}$ & em & 6376.4 & 0.06 &  & 254  & 6376.2 & 0.09 &  & 245  \\
 &  & em & 6388.6 & 0.11 &  &   & 6388.1 & 0.07 &  &   \\
 &  & em & 6390.0 & 0.07 &  &   & 6390.2 & 0.07 &  &   \\
 &  & em & 6397.4 & 0.06 &  &   & 6397.2 & 0.09 &  &   \\
 &  & em & 6407.8 & 0.06 &  &   & 6407.5 & 0.09 &  &   \\
 &  & em & 6430.2 & 0.11 &  &   & 6429.9 & 0.11 &  &   \\
\CII{} 6462.0 & $^2$F$^{\rm o}$--$^2$G & em & 6467.2 & 0.63 &  & 241 & 6466.7 & 0.36 &  & 218 \\
 &  & em & 6511.8 & 0.04 &  &   & 6511.6 & 0.07 &  &   \\
 &  & em & 6527.4 & 0.05 &  &   & 6526.9 & 0.05 &  &   \\
\fNII{} 6548.0 & $^3$P--$^1$D & em & 6553.4 & 0.12 &  & 247 & 6552.9 & 0.11 &  & 225 \\
H$\alpha$ & 2-3 & em & 6568.1 & 2.16 &  &  242 & 6567.7 & 1.84 &  & 224  \\
\CII{} 6578.1 & $^2$S--$^2$P$^{\rm o}$ & P Cyg & 6582.6 & 1.61 & --0.4 & 205 & 6582.4 & 1.74 & --0.55 & 196 \\
\CII{} 6582.9 & $^2$S--$^2$P$^{\rm o}$ & em & 6588.7 & 2.58 &  & 264  & 6588.3 & 2.28 &  & 246  \\
 \& \fNII{} 6583.5 &  \& $^3$P--$^1$D &  &  &  &  &   &  &  &  &   \\
 &  & em & 6604.1 & 0.09 &  &   & 6603.8 & 0.12 &  &   \\
 &  & em & 6623.1 & 0.15 &  &   & 6622.9 & 0.18 &  &   \\
 &  & em & 6626.9 & 0.13 &  &   & 6626.7 & 0.16 &  &   \\
 &  & em & 6673.7 & 0.07 &  &   & 6673.3 & 0.15 &  &   \\
\HeI{} 6678.2 & $^1$P$^{\rm o}$--$^1$D & P Cyg & 6682.9 & 0.61 & --0.61 & 211 & 6682.7 & 0.69 & --0.19 & 202 \\
\fSII{} 6716.4 & $^4$S$^o$--$^2$D$^{\rm o}$ & em & 6722.1 & 0.27 &  & 255 & 6721.9 & 0.33 &  & 246 \\
 &  & em & 6729.3 & 0.22 &  &   & 6729.0 & 0.23 &  &   \\
\CII{} 6727.3 & $^4$D--$^4$D$^{\rm o}$ & em & 6732.8 & 0.04 &  & 245 & 6732.2 & 0.05 &  & 219 \\
\fSII{} 6730.8 & $^4$S$^o$--$^2$D$^{\rm o}$ & em & 6736.4 & 0.19 &  & 250 & 6736.2 & 0.23 &  & 241 \\
\CII{} 6733.6 & $^4$D--$^4$D$^{\rm o}$ & em & 6738.8 & 0.05 &  & 232 & 6739.0 & 0.11 &  & 241 \\
\CII{} 6738.6 & $^4$D--$^4$D$^{\rm o}$ & em & 6744.3 & 0.07 &  & 254 & 6744.0 & 0.1 &  & 240 \\
\CII{} 6742.4 & $^4$D--$^4$D$^{\rm o}$ & em & 6748.2 & 0.09 &  & 258 & 6747.7 & 0.18 &  & 236 \\
\CII{} 6750.5 & $^4$D--$^4$D$^{\rm o}$ & em & 6756.5 & 0.1 &  & 267 & 6756.3 & 0.1 &  & 258 \\
\CII{} 6755.2 & $^4$D--$^4$D$^{\rm o}$ & em & 6760.8 & 0.04 &  & 249 & 6760.6 & 0.05 &  & 240 \\
 &  & em & 6767.4 & 0.03 &  &   &  &  &  &   \\
\CII{} 6780.6 & $^4$P$^{\rm o}$--$^4$D & P Cyg & 6785.2 & 0.15 & --0.37 & 204 & 6785.2 & 0.21 & --0.24 & 204 \\
\CII{} 6783.9 & $^4$P$^{\rm o}$--$^4$D & em & 6789.9 & 0.23 &  & 265 & 6789.8 & 0.22 &  & 261 \\
\CII{} 6787.2 & $^4$P$^{\rm o}$--$^4$D & em & 6793.2 & 0.18 &  & 265 & 6793.0 & 0.16 &  & 256 \\
\CII{} 6791.5 & $^4$P$^{\rm o}$--$^4$D & em & 6797.6 & 0.18 &  & 270 & 6797.3 & 0.15 &  & 256 \\
\CII{} 6798.1 & $^4$P$^{\rm o}$--$^4$D & em & 6803.7 & 0.09 &  & 247 & 6803.5 & 0.12 &  & 238 \\
\CII{} 6800.7 & $^4$P$^{\rm o}$--$^4$D & em & 6806.6 & 0.2 &  & 260 & 6806.4 & 0.13 &  & 251 \\
\CII{} 6812.3 & $^4$P$^{\rm o}$--$^4$D & em & 6818.1 & 0.17 &  & 255 & 6817.6 & 0.12 &  & 233 \\
 &  & em & 6827.2 & 0.11 &  &   &  &  &  &   \\
 &  & em & 6881.3 & 0.11 &  &   & 6826.8 & 0.11 &  &   \\
 &  & em & 6889.0 & 0.04 &  &   & 6831.8 & 0.03 &  &   \\
 &  & em & 6912.2 & 0.09 &  &   & 6911.7 & 0.14 &  &   \\
 &  & em & 6916.1 & 0.05 &  &   & 6915.7 & 0.11 &  &   \\
 &  & em & 6927.1 & 0.08 &  &   & 6927.0 & 0.11 &  &   \\
 &  & em & 6935.9 & 0.12 &  &   & 6935.5 & 0.14 &  &   \\
 &  & em & 7000.0 & 0.03 &  &   & 6999.8 & 0.03 &  &   \\
\CIII{} 7037.3 & $^1$P$^{\rm o}$--$^1$D & em & 7043.0 & 0.09 &  & 243 & 7042.8 & 0.05 &  & 235 \\
\CII{} 7046.3 & $^4$S--$^4$P$^{\rm o}$ & em & 7051.9 & 0.12 &  & 238 & 7051.6 & 0.11 &  & 226 \\
\CII{} 7053.1 & $^4$S--$^4$P$^{\rm o}$ & em & 7058.8 & 0.22 &  & 242 & 7058.5 & 0.2 &  & 230 \\
\HeI{} 7065.2 & $^3$P$^{\rm o}$--$^3$S & em & 7071.2 & 1.03 &  & 255 & 7070.9 & 1.23 &  & 242 \\
 &  & em & 7108.3 & 0.14 &  &   & 7108.0 & 0.1 &  &   \\
\CII{} 7112.5 & $^4$D--$^4$F$^{\rm o}$ & em & 7118.4 & 0.11 &  & 249 & 7118.1 & 0.13 &  & 236 \\
\CII{} 7115.6 & $^4$D--$^4$F$^{\rm o}$ & em & 7121.1 & 0.14 &  & 232 & 7121.0 & 0.14 &  & 228 \\
\CII{} 7119.9 & $^4$D--$^4$F$^{\rm o}$ & em & 7125.8 & 0.22 &  & 249 & 7125.6 & 0.17 &  & 240 \\
\CII{} 7125.7 & $^4$D--$^4$F$^{\rm o}$ & em & 7131.5 & 0.08 &  & 244 & 7131.2 & 0.09 &  & 232 \\
\CII{} 7134.1 & $^4$D--$^4$F$^{\rm o}$ & em & 7139.7 & 0.15 &  & 236 & 7139.2 & 0.18 &  & 215 \\
 &  & em & 7166.0 & 0.08 &  &   & 7165.7 & 0.12 &  &   \\
\CII{} 7231.3 & $^2$P$^{\rm o}$--$^2$D & em & 7237.4 & 2.9 &  & 253 & 7236.9 & 2.15 &  & 232 \\
\CII{} 7236.4 & $^2$P$^{\rm o}$--$^2$D & em & 7242.6 & 3.53 &  & 257 & 7242.3 & 3.13 &  & 245 \\
 &  & em & 7273.6 & 0.13 &  &   & 7273.1 & 0.14 &  &   \\
 &  & em & 7282.9 & 0.04 &  &   &  &  &  &   \\
\HeI{} 7281.4 & $^1$P$^{\rm o}$--$^1$S & em & 7287.4 & 0.29 &  & 247 & 7287.0 & 0.24 &  & 231 \\
\fOII{} 7320.0 & $^2$D$^{\rm o}$--$^2$P$^{\rm o}$ & em & 7325.5 & 0.11 &  & 225 & 7325.2 & 0.1 &  & 213 \\
\fOII{} 7329.7 & $^2$D$^{\rm o}$--$^2$P$^{\rm o}$ & em & 7336.2 & 0.06 &  & 266 & 7335.7 & 0.11 &  & 246 \\
 &  & em & 7376.1 & 0.05 &  &   & 7375.8 & 0.06 &  &   \\
 &  & em & 7439.5 & 0.05 &  &   & 7439.1 & 0.05 &  &   \\
 &  & em & 7465.3 & 0.02 &  &   &  &  &  &   \\
 &  & em & 7472.1 & 0.01 &  &   &  &  &  &   \\
 &  & em & 7475.5 & 0.08 &  &   & 7475.2 & 0.11 &  &   \\
\CII{} 7505.3 & $^2$P$^{\rm o}$--$^2$D & em & 7511.3 & 0.21 &  & 240 & 7510.8 & 0.18 &  & 220 \\
\CII{} 7508.9 & $^2$P$^{\rm o}$--$^2$P & em & 7515.6 & 0.21 &  & 268 & 7515.1 & 0.2 &  & 248 \\
\CII{} 7520 & $^2$P$^{\rm o}$--$^2$P & em & 7525.9 & 0.27 &  & 235 & 7525.6 & 0.3 &  & 223 \\
 &  & em & 7530.0 & 0.05 &  &   & 7529.9 & 0.06 &  &   \\
\CII{} 7530.6 & $^2$P$^{\rm o}$--$^2$P & em & 7536.8 & 0.08 &  & 247 & 7536.3 & 0.08 &  & 227 \\
 &  & em & 7571.0 & 0.03 &  &   &  &  &  &   \\
 &  & em & 7575.3 & 0.04 &  &   &  &  &  &   \\
 &  & em & 7580.3 & 0.12 &  &   & 7580.0 & 0.14 &  &   \\
\OI{} 7771.9 & $^5$S$^{\rm o}$-$^5$P & abs & 7776.6 & 0.15 &  & 181 & 7776.6 &  & --0.5 & 181 \\
\OI{} 7775.4 & $^5$S$^{\rm o}$-$^5$P & P Cyg & 7780.7 & 0.17 &  & 205 & 7780.7 & 0.2 & --0.28 & 205 \\
 &  & em & 7802.3 & 0.06 &  &   & 7802.0 & 0.04 &  &   \\
 &  & em & 7822.2 & 0.04 &  &   &  &  &  &   \\
 &  & em & 7867.1 & 0.13 &  &   & 7866.8 & 0.16 &  &   \\
 &  & em & 7883.4 & 0.05 &  &   & 7883.0 & 0.11 &  &   \\
 &  & em & 7902.7 & 0.07 &  &   & 7902.4 & 0.14 &  &   \\
\CII{} 8037.7 & $^4$P--$^4$P$^{\rm o}$ & em & 8045.0 & 0.1 &  & 273 & 8044.6 & 0.13 &  & 258 \\
\CII{} 8062.1 & $^4$P--$^4$P$^{\rm o}$ & em & 8068.9 & 0.12 &  & 253 & 8068.5 & 0.14 &  & 238 \\
\CII{} 8076.6 & $^4$P--$^4$P$^{\rm o}$ & em & 8083.1 & 0.13 &  & 241 & 8082.7 & 0.11 &  & 227 \\
 &  & em & 8146.7 & 0.11 &  &   & 8146.5 & 0.17 &  &   \\
 &  & abs & 8169.8 &  & --0.44 &   & 8169.6 &  & --0.35 &   \\
 &  & abs & 8178.2 &  & --0.61 &   & 8178.0 &  & --0.41 &   \\
\CII{} 8214.5 & $^2$G--$^2$H$^{\rm o}$ & em & 8221.4 & 0.13 &  & 252 & 8221.0 & 0.2 &  & 237 \\
 &  & em & 8240.6 & 0.05 &  &   & 8240.6 & 0.05 &  &   \\
 &  & abs & 8288.7 &  & --0.32 &   & 8288.0 &  & --0.28 &   \\
 &  & abs & 8294.6 &  & --0.51 &   & 8294.2 &  & --0.18 &   \\
 &  & em & 8300.2 & 0.31 &  &   & 8299.8 & 0.23 &  &   \\
 &  & em & 8341.9 & 0.03 &  &   &  &  &  &   \\
 &  & em & 8345.8 & 0.08 &  &   & 8345.4 & 0.06 &  &   \\
 &  & em & 8357.1 & 0.09 &  &   & 8356.6 & 0.13 &  &   \\
 &  & em & 8399.4 & 0.03 &  &   &  &  &  &   \\
 &  & em & 8417.0 & 0.02 &  &   &  &  &  &   \\
 &  & em & 8420.1 & 0.04 &  &   &  &  &  &   \\
 &  & em & 8430.4 & 0.01 &  &   &  &  &  &   \\
\OI{} 8446.4 & $^3$S$^{\rm o}$--$^3$P & em & 8453.1 & 0.16 &  & 238 & 8452.8 & 0.28 &  & 227 \\
\CII{} 8682.5 & $^2$S--$^2$P$^{\rm o}$ & em & 8689.9 & 0.53 &  & 256 & 8689.5 & 0.6 &  & 242 \\
\CII{} 8696.7 & $^2$S--$^2$P$^{\rm o}$ & em & 8704.1 & 0.3 &  & 255 & 8703.7 & 0.36 &  & 242 \\
 &  & em & 8801.8 & 0.45 &  &   &  &  &  &   \\
 &  & em & 8808.1 & 0.24 &  &   &  &  &  &   \\
 &  & em & 8827.4 & 0.07 &  &   &  &  &  &   \\
 &  & em & 8880.8 & 0.16 &  &   & 8880.2 & 0.15 &  &   \\
 &  & em & 8919.9 & 0.07 &  &   &  &  &  &   \\
 &  & em & 8958.8 & 0.1 &  &   &  &  &  &   \\
\CII{} 9089.1 & $^2$G--$^2$H$^{\rm o}$ & em & 9096.6 & 0.06 &  & 248 & 9095.8 & 0.07 &  & 221 \\
 &  & em & 9102.8 & 0.08 &  &   &  &  &  &   \\
 &  & em & 9225.7 & 0.09 &  &   & 9225.2 & 0.17 &  &   \\
 &  & em & 9231.0 & 0.09 &  &   & 9230.6 & 0.12 &  &   \\
\CII{} 9229.6 & $^2$D--$^2$F$^{\rm o}$ & em & 9237.1 & 0.43 &  & 244 & 9236.9 & 0.44 &  & 237 \\
 &  & em & 9243.8 & 0.17 &  &   & 9243.7 & 0.23 &  &   \\
 &  & em & 9251.8 & 0.07 &  &   & 9251.3 & 0.11 &  &   \\
 &  & em & 9257.5 & 0.09 &  &   & 9257.3 & 0.1 &  &   \\
 &  & em &  &  &  &   & 9412.8 & 0.19 &  &   \\
 &  & em &  &  &  &   & 9857.6 & 0.26 &  &   \\
\CII{} 9904 & $^2$F$^{\rm o}$--$^2$G & em &  &  &  &   & 9911.0 & 0.65 &  & 212 \\
 &  & em &  &  &  &   & 10245.0 & 0.2 &  &   \\
 &  & em &  &  &  &   & 10264.0 & 0.13 &  &   \\
\HeI{} 10830 & $^3$S--$^3$P$^{\rm o}$ & em &  &  &  &   & 10838.9 & 0.59 &  & 247 
\enddata
\tablenotetext{a}{2017 H$\beta$ + \CII~blend flux is 5.0 $\times 10^{-15}$ erg s$^{-1}$ cm$^{-2}$ \AA$^{-1}$}
\tablenotetext{b}{2019 H$\beta$ + \CII~blend flux is 5.7 $\times 10^{-15}$ erg s$^{-1}$ cm$^{-2}$ \AA$^{-1}$}
\tablenotetext{c}{All quoted radial velocities are heliocentric, positive values.}
\tablecomments{Table 1 is published in its entirety in the machine readable format.  A portion is shown here for guidance regarding its form and content.}\label{tab:Jmeas}
\end{deluxetable*}
\clearpage

\begin{deluxetable*}{ccccc}
\tablecaption{\CII{} Recombination Intensities\label{tab:recomb}}
\tablecolumns{5}
\tablenum{2}
\tablewidth{0pt}
\tablehead{
\colhead{Line (Doublets)} &
\colhead{RR Recombination Coeffs\tablenotemark{a}} &
\colhead{Predicted Recomb Intensity} & \multicolumn{2}{c}{Observed Intensity} \\
\colhead{(\AA)} & \colhead{(10$^{-14}$ cm$^{3}$s$^{-1}$)} &
\colhead{(F$_{4267}=100$)} & \colhead{J0608} & \colhead{IC 418\tablenotemark{c}}
}
\startdata
\CII{} $\lambda$4267 & 27.6 & 100.0 & 100.0 & 100 \\
\CII{} $\lambda$5342 & 1.84 & 5.3 & 10.7 & 4.9 \\
\CII{} $\lambda$6151 & 1.69 & 4.3 & 13.4 & 4.4 \\
\CII{} $\lambda$6462 & 4.29 & 10.3 & 13.8 & 10.2 \\
\CII{} $\lambda$9230 & 3.65 & 6.1 & 10.7 & \nodata \\
\hline
\colhead{Line (Multiplets)} &\colhead{DR Recombination Coeffs\tablenotemark{b}} &\colhead{} & \colhead{} &\colhead{}\\
\hline
\CII{} 6095-6103 (24) & 0.17 & 0.43 & 39.0 & 0.19 \\
\CII{} 4372-4374 (45) & 0.99 & 3.5 & 18.7 & \nodata \\
\CII{} 7112-7134 (20) & 1.33 & 2.9 & 27.0 & 2.9 \\
\enddata
\tablenotetext{a}{Recombination coefficients are for $T_{\rm eff} =10^4$ K (Davey et al.\ 2000)}
\tablenotetext{b}{Dielectronic recombination coefficients are for $T_{\rm eff} =10^4$ K (Sochi \& Storey 2013)}
\tablenotetext{c}{IC 418 intensities from Sharpee et al.\ (2003)}
\end{deluxetable*}

\acknowledgments

We thank G. Ferland, G. Jacoby, P. Massey, M. Phillips, A. Pradhan, I. Soszynski, R. van der Marel, and K.~Werner for useful discussions and data, and are grateful to N. Hambly for supplying the SuperCOSMOS images. We appreciate comments by the anonymous referee, which have improved this manuscript. This work has made use of data from the European Space Agency (ESA) mission \Gaia{} (\url{https://www.cosmos.esa.int/gaia}), processed by the \Gaia{} Data Processing and Analysis Consortium (DPAC, \url{https://www.cosmos.esa.int/web/gaia/dpac/consortium}). Funding for the DPAC has been provided by national institutions, in particular the institutions participating in the \Gaia{} Multilateral Agreement.



\end{document}